\documentclass{aa}  
\bibpunct{(}{)}{;}{a}{}{,}
\usepackage{graphicx}
\usepackage{soul}
\usepackage{txfonts}
\usepackage{lscape}
\usepackage[draft=true]{hyperref}
\usepackage{tabularx}       
\usepackage{longtable}
\usepackage{color}

\newcommand{\degree}{\ensuremath{^\circ}}
\newcommand{\px}{\ensuremath{P_X}}
\newcommand{\pq}{\ensuremath{P_Q}}
\newcommand{\pu}{\ensuremath{P_U}}

\begin{document} 

\title{Spectral and Temporal Variability of Earth Observed in Polarization}

\author{ Michael F. Sterzik\inst{1}
  \and
  Stefano Bagnulo\inst{2}
 \and
  Daphne M. Stam\inst{3}
 \and
 Claudia Emde\inst{4}
  \and
 Mihail Manev\inst{4}
 }

\institute{
 European Southern Observatory, Karl-Schwarzschild-Str. 2, D-85748 Garching, Germany 
 \email{msterzik@eso.org}
  \and
  Armagh Observatory and Planetarium, College Hill, Armagh BT61 9DG, UK
\and
Faculty of Aerospace Engineering, Delft University of Technology, Kluyverweg 1, 2629 HS Delft, The Netherlands
\and
Meteorological Institute, Ludwig-Maximilians-University,
  Theresienstr. 37, D-80333 Munich, Germany
  }

\date{Received 9-Sep-2018; Accepted 28-Nov-2018}


\abstract
{Earthshine, i.e. sun-light scattered by Earth and back-reflected from the lunar surface to Earth, allows
observations of Earth's total flux and polarization with ground-based astronomical facilities on timescales from minutes to years. 
Like flux spectra, polarization spectra exhibit imprints of Earth's atmospheric and surface properties. 
Earth's polarization spectra may prove an important  benchmark to constrain expected bio-signatures of 
Earth-like planets observed with future telescopes.
} 
{
We derive Earth's polarimetric phase curve from a statistically significant sample of Earthshine polarization spectra.  
The impact of changing Earth views on the variation of polarization spectra is investigated.
}
{
We present a comprehensive set of  spectropolarimetric observations of Earthshine as obtained by FORS2 at the VLT for phase angles from 50\degree\ to 135\degree\ (Sun--Earth--Moon angle), covering a spectral range from 4300\AA\ to 9200\AA. The degree of polarization in $B, V, R, I$ passbands, the differential polarization vegetation index, and the equivalent width of  the O$_2$-A polarization band around 7600\AA \, are determined with absolute errors around 0.1\% in the degree of polarization.
Earthshine polarization spectra are corrected for the effect of depolarization introduced by backscattering on the lunar surface, introducing systematic errors of the order of 1\% in the degree of polarization.
}
{
Distinct viewing sceneries such as observing the Atlantic or Pacific side in Earthshine yield statistically different phase curves. 
The equivalent width defined for the O$_2$-A band polarization is found to vary from -50\AA\ to +20\AA. A differential polarized vegetation index is introduced and reveals a larger vegetation signal for those viewing sceneries that contain  larger fractions of vegetated surface areas. We corroborate the observed correlations with theoretical models from the literature, and conclude that the Vegetation Red Edge (VRE) is a robust and sensitive signature in polarization spectra of planet Earth.}
{
The overall behaviour of polarization of planet Earth in the continuum and in the O$_2$-A band can be explained by existing models. Bio-signatures such as the O$_2$-A band and the VRE are detectable in Earthshine polarization with a high degree of significance and sensitivity.
An in-depth understanding of Earthshine's temporal and spectral variability requires improved models  of Earth's biosphere, as a pre-requisite to interpret possible detections of polarised bio-signatures in Earth-like exoplanets in the future.}

\keywords{Astrobiology, Earth, Polarization, Scattering}

\maketitle
%

\section{Introduction}

With the discovery of the first potentially habitable planet in the solar neighborhood by \citet{AngladaEscude:2016hq}, 
the quest for remote detection of signatures of life in other worlds has become one of the most exigent problems in modern astrophysics.  
Remotely detectable signs of life are being carefully assessed \citep[see the reviews of][]{Schwieterman:2018df, Catling:2018dt}, and observational prospects and requirements are being mapped onto the next generation of astronomical facilities  \citep[see the review of][]{Fujii:2018fc}. 
Facing colossal difficulties, further progress is needed in all areas encompassing astrobiology, theoretical concepts and frameworks, methodology, models and instrumental capabilities.

For the time being, Earth is the only inhabited planet we know.  Planet Earth is therefore a benchmark object to infer biosignatures of life as we know it today. The signature of Earth seen as an exoplanet, thus from afar, depends strongly on the local illumination and viewing geometries and cannot be derived from Earth remote-sensing observations such as taken by low-Earth-orbit satellites. One way to study Earth from afar is to observe Earthshine. Earthshine is sunlight scattered by Earth and reflected from the lunar surface back to Earth, where it can be observed with ground-based astronomical facilities \citep[see the reviews of][]{Arnold:2008ww, Vazquez:2010it}.  Biosignatures such as the Vegetation Red Edge (VRE) and trace gases of biotic origin such as O$_2$ and CH$_4$ can be inferred from spectroscopic observations \citep{2006ApJ...644..551T}. Different surface sceneries of Earth are probed as the Sun--Earth--Moon phase angle $\alpha$ and corresponding viewing geometry changes. Temporal changes in the associated light curve contain further information about surface and atmospheric properties of Earth, mainly through spatial and temporal modulation of Earth's albedo. For example, high cadence photometric time series measurements from the {\sl Deep Space Climate Observatory} imaging Earth from afar can be used to reconstruct planetary rotation, cloud patterns, and surface types without prior knowledge of its properties in detail \citep{Jiang:2018dw}. 

However, detection and reliable extraction of bio-signatures from Earth's reflectivity spectra remains difficult. In the case of Earthshine, the light's final transmission through Earth's atmosphere contaminates the signatures in the Earthshine's spectral flux through extinction by scattering and absorption, and has to be carefully corrected for. 
Extinction by Earth's atmosphere does, however, usually not change the degree of polarization $P$ of light coming from an astronomical source, as both polarized and unpolarized fluxes are affected to the same proportion. Polarimetry implemented as an intrinsically differential measurement technique should therefore overcome most telluric calibration issues and allow more reliable measurements from the ground.  Polarization spectra, in general, also have a higher diagnostic value to fully characterize the scattering and reflecting particles and surfaces than flux spectra. Therefore we have explored spectropolarimetry of Earthshine at optical wavelengths to improve the extraction of  characteristic properties of Earth contained in the sunlight that Earth reflects \citep{Sterzik:2012gk}. The method allowed to determine the fractional contribution of clouds and ocean surfaces, and could distinguish visible areas of vegetation as small as 10\% by comparing two datasets from different epochs with different aspects of Earth. While the measured O$_2$-A band strength and the VRE feature were fully compatible with the results from spectral polarization models for Earth-type exoplanets  \citep{Stam:2008ij}, the shape of the polarized spectral continuum remained unexplained: it is significantly flatter in the red spectral part than expected from the models.  

Relatively flat polarization spectra of Earthshine were corroborated by \citet{2013PASJ...65...38T}, who presented a series of polarization spectra covering phase angles between 49\degree \; and 96\degree \; that were recorded in 5 consecutive nights. \citet{2014A&A...562L...5M} extended the wavelength regime for spectropolarimetry of Earthshine to the near-infrared, and found particularly high degrees of polarization for the H$_2$O-band around 1.12$\mu$m.   

Classic observations of Earthshine polarization were presented already by \citet{1957SAnAp...4....3D}, covering phases from around 30\degree \; to 140\degree \; in the $V$-band. The main characteristics of the polarization phase curve of Earth were qualitatively and quantitatively explained.  \citet{1957SAnAp...4....3D} found a steady increase in the fractional polarization of Earthshine from about $P\approx$2\% around a phase of 30\degree, to around 10\% peek values for 100\degree, and a decrease for larger phase angles. 
\citet{1957SAnAp...4....3D}  noted a wavelength dependance of depolarization caused by backscattering at the lunar surface and  concluded that fractional polarization of the light scattered by Earth is around $P_V \approx$33\% around quadrature.

Because Earthshine is reflected by the lunar surface, ground-based observations have to be corrected for the resulting change of the degree of polarization. Indeed, \citet{2013A&A...556A.117B} probed Earthshine polarization in four $B, V, R, I$  photometric bandpasses for phase angles between 30\degree \; and 110\degree. They demonstrated that the reflection by different regions on the Moon with different albedo's (Highlands and Mare) has a significant and distinct impact on the polarization measured in Earthshine. To mitigate this problem, they introduced a method to correct for lunar depolarization efficiency that depends on wavelength and lunar albedo. Their measurements convincingly demonstrated that the {\sl lunar surface albedo}  needs to be taken into account when calibrating and interpreting the absolute values of Earth's degree of polarization derived from Earthshine, and that lunar albedo effects may be of the same order as those caused by variations of the Earth's albedo. 

In the last decade, a substantial amount of theoretical work has been carried out to model and simulate the polarization spectrum of Earth.  \citet{Stam:2008ij} employed a method to approximate an inhomogenuous surface albedo and clouds by using weighted sums of light reflected by horizontally homogenous planets with a specific surface reflection function (Fresnel or Lambert), covered by an anisotropic Rayleigh scattering atmosphere 
containing Mie-Scattering clouds with a fixed  optical depth (equals to ten).
These calculations covered a wide range of phase angles and a wide range of surface conditions. Results can easily be retrieved from look-up tables and enable to systematically explore a multi-parameter problem.  \citet{Karalidi:2012kx} extended the methodology to horizontally inhomogeneous clouds and surfaces of Earth-like exoplanets. \citet{Karalidi:2011ch, Karalidi:2012fc} further incorporated effects of liquid and ice water clouds on the degree of polarization and their phase dependance. An alternative Monte-Carlo approach simulates a whole Earth-type planet in full spherical geometry, and was presented by \citet{GarciaMunoz:2014hn, 2015IJAsB..14..379G}. 

The 3D Monte Carlo vector radiative transfer code MYSTIC \citep{Mayer:2009hl, Emde:2010vi} has been extended to fully spherical geometry, so that it allows to simulate Earthshine polarization spectra \citep{Emde:2017ee}. MYSTIC is one of the solvers included in the libRadtran package \citep[\url{www.libradtran.org},][]{Emde:2016eo}, which is widely used for Earth remote sensing applications. The influence of aerosols, clouds and the potential impact of sunglint was studied to explain in detail their impact on two spectra published by \citet{Sterzik:2012gk}. They incorporated three-dimensional fields of cloud properties (cloud cover, liquid and ice water content) from the ECMWF IFS weather forecast model (\url{www.ecmwf.int}) at the time of observations and a two-dimensional surface albedo map derived from MODIS satellite observations, and simulated high spatial and high spectral resolution maps of the Stokes vectors across Earth's surface. The simulations approximate the observed polarization spectra much better than the weighted sum spectra of the horizontally homogeneous planets \citep[][]{Sterzik:2012gk}.

The purpose of this paper is to extend the analysis of Earthshine polarization spectra to a much larger sample of high-quality observations. In a monitoring campaign, around 50 polarization spectra of Earthshine have been recorded between April 2011 and February 2013, all of them observed with FORS2 at the VLT. They cover phase angles of Earth between 50\degree\ and 135\degree. Suitable parameters derived from these spectra allow statistical analysis of the temporal and spectral variation of the degree of polarization. In this paper, we discuss 34 spectra of the highest quality.
One goal of this work is to establish a set of observables from Earthshine spectropolarimetry that can be used to characterize the Earth's surface and atmosphere at the time of observations, minimizing the uncertainties caused in particular by lunar (de-)polarization efficiencies.   We will extensively use the models of \citet{Stam:2008ij}  to compare them with the statistical properties in our sample and infer global trends. However, we do not intend to simulate and explain specific and individual datasets, owing to the fact that they require dedicated modeling, which is beyond the scope of this paper.

The process of acquiring and reducing our datasets is a challenging task: essentially, it consists of spectropolarimetric measurements of a fraction of the visible lunar disk not illuminated by the sun. Observationally, the lunar disk is an extended target that is  moving at a quickly changing, non--sidereal velocity, that cannot automatically be tracked by the VLT.
In addition, the Earthshine signal that is reflected by the dark fraction of the lunar disk is contaminated by  the (polarized) signal of the Moonshine (the part of the lunar disk that is illuminated by the sun).
This contamination increases with increasing lunar phase.

This paper is organized as follows. In Sects.~\ref{Sect_Observations} and \ref{Sect_Data_Reduction} we explain the procedure to obtain accurate data. In Sect.~\ref{Results} we elaborate on our results. We present our approach to correct the Earthshine polarization for lunar depolarization, and derive the polarization phase curve of Earth in $B, V, R, I$ spectral bandpasses.  We derive polarization color ratios, the differential polarized vegetation index and the equivalent width of the O$_2$-A band for our observations, look into the short-term variability of these parameters, and compare them with literature and model results. Finally, we summarize our conclusions and outlook in Sect.~\ref{Discussion}.
 

\section{Observations}\label{Sect_Observations}
In this Section we describe the instrument and instrument settings for which we have obtained Earthshine measurements, as well as the special procedures that we adopted both for data acquisition and for data reduction.

\subsection{Instrument and instrument settings}

All our observations were obtained with the FORS2 instrument \citep{1998Msngr..94....1A} of the ESO VLT.
FORS2 is a multipurpose instrument capable of imaging and low resolution spectroscopy, equipped with polarimetric optics. Polarimetric optics consist of a retarder waveplate (either a $\lambda/2$ retarder waveplate for linear polarization measurements, or a $\lambda/4$ retarder waveplate for circular polarization measurements) followed by a Wollaston prism, that splits the radiation into two beams linearly polarized in perpendicular directions, separated by 22\arcsec. In front of the retarder waveplate, a Wollaston mask with nine 22\arcsec\ slitlets prevents the superposition of the field of view with the beam split by the Wollaston prism, following the optical scheme of \citet{1983MNRAS.204.1163S}.

FORS2 allows us to directly measure the (wavelength dependent) quantities $\pq=Q/I$ and $\pu=U/I$, and thus the fractional polarization $P(\lambda)$ (or degree of polarization, in short "Polarization", in percent) defined as 
\begin{equation}
P = \sqrt{(\pq^2 + \pu^2)}.
\label{eq:pol}
\end{equation} 
The angle of polarization ($\phi$) may be obtained from
\begin{equation}
\tan( 2 \cdot \phi) = {U/Q}.
\label{eq:ang} 
\end{equation}
As our slit direction (and therefore the FORS2 instrument rotation) was  consistently oriented in east-west direction, we rotate the measured quantities $Q/I$ and $U/I$ by 90\degree\  into the instrument reference system to yield an equatorial reference system  pointing to the north celestial pole.
In the following, we calculate and discuss both  quantities $P$ and $\phi$.

Most of our observations were obtained with grism 300V, which spans the optical range between 3600\;\AA\; and 9200\;\AA. An order separating filter may be inserted in the optical beam to cut the radiation shortward 4200\;\AA\; and prevents second order contamination at $\lambda \ga 6500$\;\AA. We obtained observations with and without order separating filter and confirm that polarization spectra are not affected by contamination from second order \citep{Patat:2010hm}. We also used grism 600\,I that covers the spectral range 6700\;\AA\; to 9300\;\AA. We adopted a 2\arcsec\ slit width, for a spectral resolution of 220 and 750 with grism 300V and grism 600I, respectively. 

The performance and general calibration techniques of the polarimetric modes of the FORS2 instrument are well known and documented. In particular, the variation of the instrument polarization over the field of view, the chromatism of the retarder waveplate and the cross-talk between different  Stokes parameters has been characterized in detail \citep[see e.g.][]{2009PASP..121..993B, 2011ASPC..449...76B}. The systematic effects are typically of the order of 0.1\% or less in the field center,  which is practically negligible as contribution to systematic errors in our context.

\subsection{Data acquisition}\label{DA}

\begin{figure}
\resizebox{\hsize}{!}{\includegraphics[]{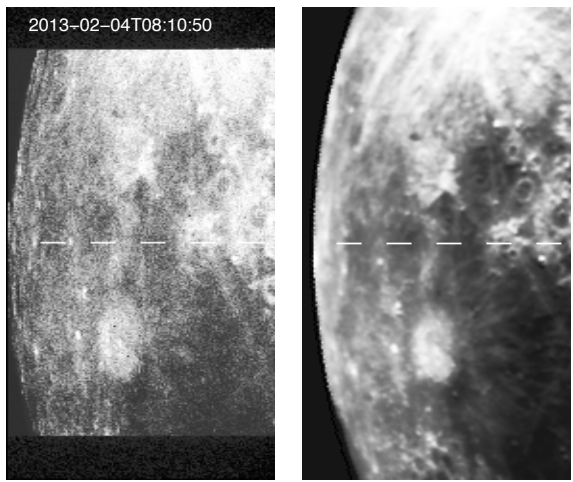}}
\caption{Left: Acquisition image of the lunar limb that contains Earthshine observed on 2013-02-04UT08:10:50. Right: map of the apparent albedo from \citet{Velikodsky:2011dg} of the same region that has been scaled and rotated. Corresponding FORS2 slitlets are superposed on both images.  }
\label{Fig:acq}
\end{figure}

The FORS2 detector consists of two chips separated by a 4\arcsec\ gap. As a general rule, we aimed at pointing to the Moon with chip 1 containing Earthshine and chip 2 background sky, hence with instrument position angle 
on sky (relative to the north celestial pole) equal to 90\degree\ when pointing to the waxing moon, or 270\degree\ when pointing to the waning moon. Acquisition consisted of a first pointing to the centre of the Moon followed by a 15\arcmin\ offset either to the East or to the West, and a final tuning using more direct imaging at the lunar limb and through-slit images. 
An example of a typical acquisition image is shown in Fig.~\ref{Fig:acq}, on the left, obtained immediately before the dataset corresponding to ID F.6 (Tab.~\ref{Tab:Log}). 
The lunar limb is clearly seen, and five 22\arcsec\ long slitlets on chip 1 are superposed to scale on the image to indicate their actual position, because through-slit images do not allow to reliably identify the actual position on the Moon.  We always tried to position the limb in the gap between both chips, to avoid observing the immediate vicinity of the limb. The four slitlets of chip 1 that cover the background are not shown in the image.

Science observations were generally obtained with the retarder waveplate set at all position angles $0,22.5,45,\dots,337.5\degr$, in order to apply the so called ``beam swapping technique'' \citep[see, e.g.,][]{2009PASP..121..993B}. Basically, at each position angle of the retarder waveplate, the light flux is measured in perpendicular polarization states. The measurement is repeated after a 45\degr\ rotation of the $\lambda/2$ retarder waveplate, thus with swapped polarization states in the two beams split by the Wollaston prism. The combination of the fluxes measured in the various beams allows one to remove most of the instrumental effects and calculate the reduced Stokes parameters $\px = X/I$ (where $X$ represents the various Stokes parameters $Q$, $U$, and $I$ is the unpolarized intensity) with an uncertainty ideally given by $(S/N)^{-1}$. Practically, the actual uncertainty of our Earthshine measurements is not determined by the photon-noise, but is dominated by systematic instrumental effects and by the fact that the reflection properties of the lunar surface are only  approximately known (see Sect.~\ref{depol} below). Uncertainties were also statistically checked through inspections of the so-called null profiles, as discussed by \citet{2009PASP..121..993B}.

\subsection{Observations and Earth Viewing Geometry}

In Tab.~\ref{Tab:Log} we list all parameters that characterize a given cycle of our polarimetric observations of Earthshine:
a unique identifier (ID), the date and time when the observation cycle started, the airmass at this time, and the instrumental set-up with the grism used. We also list the exposure times of individual exposures. The total exposure time is obtained by multiplying this number by the total number of exposures at individual settings of the retarder plate, which is also specified in the table. The phase angle $\alpha$ is defined as the angle between Sun -- Earth -- Moon. The viewing sceneries of 
Earth refer either to observations of the region to the west of the place of the observations in 
Chile, i.e. the Pacific ("P", observed after sunset), or to observations of the region to the
east, i.e. the Atlantic ("A", observed before sunrise). 
As Earth rotates around its axis, its phase angle changes slowly.
Also global weather patterns are changing slowly, but continuously. We do expect them to modulate 
the observed polarization steadily, but not abruptly (see Sect.~\ref{STV}). We have indicated observations that belong to the same sequence of observations by the same capital letter for their ID in Tab.~\ref{Tab:Log} and \ref{Tab:PEarth}, while distinguishing individual observations with a numerical suffix. Observations a few weeks later, however, 
are likely influenced by completely different cloud coverage maps, that may have a strong effect 
on the overall polarization signal. Distinct observing sequences are therefore separated by larger horizontal spaces in the table and have different letters in their IDs.

\begin{figure*}
\centering
\includegraphics[width=19cm]{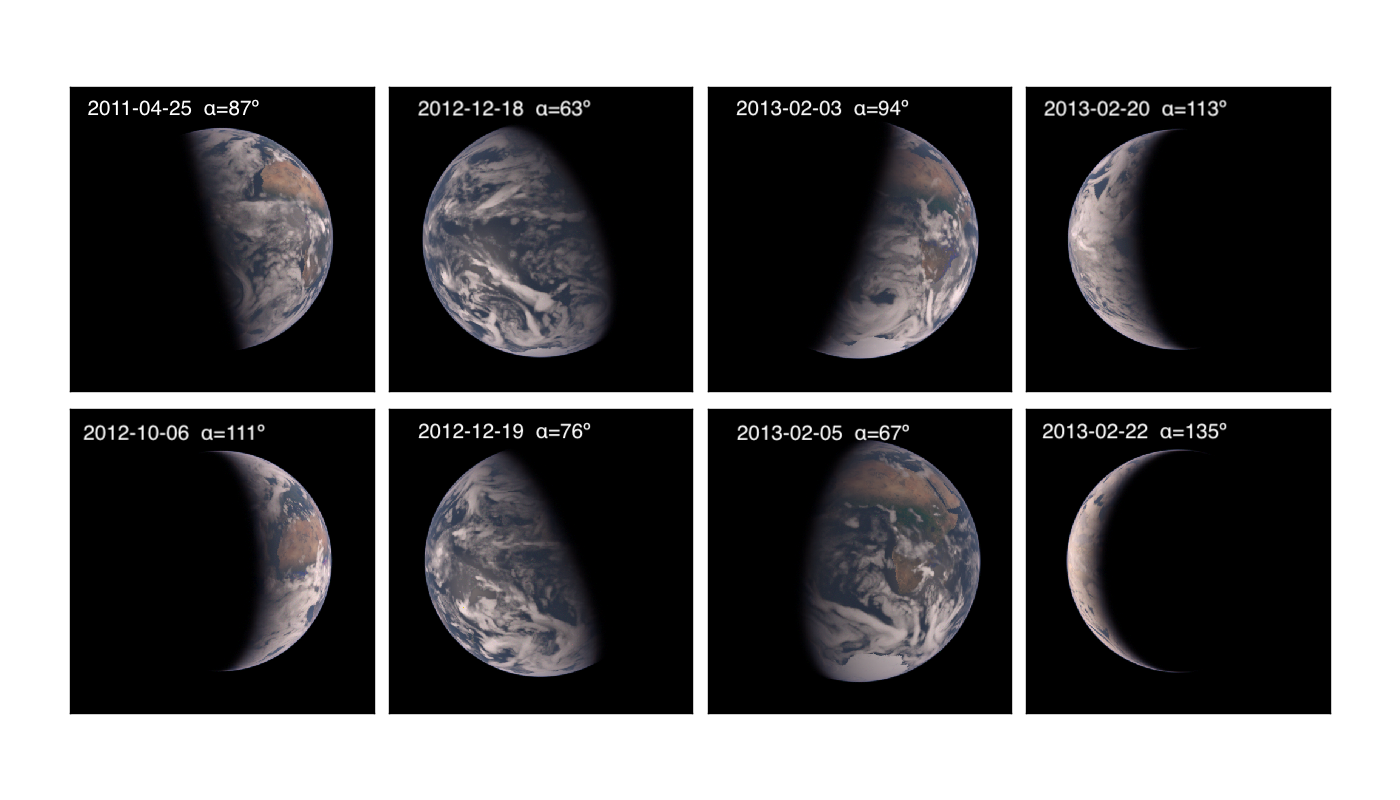} 
\caption{The aspect of the Earth for eight representative observing epochs. True color RGB composite images of Earth as seen from the Moon, simulated using the MYSTIC radiative transfer model with cloud data from the ECMWF forecast model and MODIS surface albedo. The epochs correspond to the phase angles during observations.}
\label{Fig:8Earth}
\end{figure*}

A few examples of Earth's appearance during the observations for representative observing epochs are shown in Fig.~\ref{Fig:8Earth} as true color RGB composite images. The images were generated by MYSTIC radiative transfer simulations at wavelengths of 645\,nm (red), 555\,nm (green) and 469\,nm (blue) as described in \citet{Emde:2017ee}. As input we have used three-dimensional cloud field data from the ECMWF model closest to the date and time of the observations and land surface albedo data derived from MODIS \citep{Schaaf:2002}. The ocean surface is simulated using a polarized bidirectional reflectance distribution function \citep{tsang1985, mishchenko1997}. The aspect of Earth's surface is seen from the lunar center. 

\section{Data Reduction}\label{Sect_Data_Reduction}
The FORS pipeline  \citep{Izzo:2010jr} handles spectropolarimeric data of point sources, but is not designed for spectropolarimetry of extended objects. Therefore various dedicated procedures were developed and applied to our Earthshine data.

\subsection{Frame preprocessing}
All science frames were pre-processed with the ESO FORS pipeline (vers. 4.8.7.) to remove bias and perform a 2D wavelength calibration. Among the various final pipeline products we used only the frames mapped in wavelength and corrected for field distorsion, but not flatfielded.

\subsection{Flat fielding}
In contrast to what is normally expected for polarimetric data obtained with the beam-swapping technique, flat-fielding is important in the reduction of spectropolarimetry of Earthshine. The reason is that Moonshine background is not constant but must be interpolated in chip 2 and then (linearly) extrapolated to chip1 to be correctly substracted from Earthshine observations (see Sect.~\ref{Sect_Background}). Flatfield spectropolarimetric images may be obtained using either the calibration unit, or using twilight sky. Twilight sky data should be more suitable than data obtained with the internal calibration unit, because twilight sky images follow the same optical path as science data, and therefore should better represent large-scale spatial gradients in the system response.  Ideally, flat-field data should be obtained through a continuously rotating retarder waveplate as to measure a totally non polarized image. This option is not implemented by the instrument software, therefore our master flat-fields were obtained adding up images obtained at various retarder waveplate position angles. Data were smoothed in wavelength, and the master flatfield was normalised along the direction perpendicular to the dispersion, by dividing the entire image by a 1D spectrum obtained as average of its central 50 raws. 
We note that with a suitable combination of the frames obtained at various positions 
of the retarder waveplate (e.g., all positions $0\degr, 22.5\degr, 45\degr, \ldots, 337.5\degr$) 
one could in theory obtain a totally unpolarised master flat field, but in practise this did 
not work, because the sky polarization changes quickly during twilight. This appeared 
not to be a major problem, because what needs to be calibrated to correct for field distortion,
is the total flux, which is obtained by adding the two beams of the master-flat field. 

Sky flatfields were available for datasets with IDs D.x and observations later than December 2012, whereas screen flatfields were available for all datasets. We used sky flatfields  whenever available.  In order to minimize the effect of using different type of flatfields,  we constructed stacked and smoothed sky flatfields from those available, and applied the  correction to all screen flatfields. We compared the extracted spectra for pure sky- and screen-corrected flat field calibration, and did not find any significant differences in the extracted spectra. As we are not aware of changes of instrument configuration happening during April 2011 to December 2012 that could affect the optical beam, we are confident that this procedure ensures an optimal flatfield correction even for those datasets without corresponding twilight sky calibration data available. 

\subsection{Background subtraction}\label{Sect_Background}

Background subtraction is a crucial step in the entire data reduction process. Its accuracy relies on the assumption that (1) field distortions have been correctly removed by 2D wavelength calibration and flatfielding, (2) that Moonshine intensity decreases linearly with the distance from the terminator, and (3) that Moonshine polarisation is constant within the 7\arcmin\ range of the PMOS slit. For each wavelength bin, at the spatial position $y$, the Earthshine polarisation spectrum is then obtained as

\begin{equation}
  \px^{\rm (ES)}(\lambda,y) = \frac{\px^{\rm (tot)}(\lambda,y) F^{\rm (tot)}(\lambda,y) - \px^{\rm (bkg)}(\lambda) F^{\rm (bkg)}(\lambda,y)}{ F^{\rm (tot)}(\lambda,y) -  F^{\rm (bkg)}(\lambda,y)}
\end{equation}
where
$\px^{\rm (tot)}(\lambda,y)$ is the polarisation measured at wavelength $\lambda$ and position $y$ on the CCD (e.g., measured in arcecs from the edge of the CCD that is outside the lunar disk);
$F^{\rm (tot)}(\lambda,y)$ is the total measured intensity;
$\px^{\rm (bkg)}(\lambda)$ is the polarisation of the background at wavelength $\lambda$ and average over the $y$ range in chip 2;
$F^{\rm (bkg)}(\lambda,y)$ is obtained as linear extrapolation of the flux measured in chip 2. The background subtraction method also works in absorption band regions (like those around  the O$_2$ bands), although overall less photons are available as compared to the continuum, and a corresponding reduction of the $S/N$ ratio achieved.

In the ideal case, $\px^{\rm (ES)}(\lambda,y)$ should actually be independent of $y$ (i.e. along the slit direction), but this is rarely the case. This is either because instrument distortions were not perfectly corrected, or the background does not change exactly linearly with distance. However, the largest contribution is due to lunar albedo changes along the slit (see Fig.~\ref{Fig:acq}), and its effect on the lunar polarization efficiency, as discussed in Sect.~\ref{depol}.  Indeed, a clear anti-correlation between intensity and fractional polarization exists along each slit. 
Therefore we always average the degree of polarization along the slit. 

\section{Results}\label{Results}

\subsection{Spectral characteristics}

\begin{figure*}
\centering
\includegraphics[width=20cm, trim = 3cm 1cm 0 0]{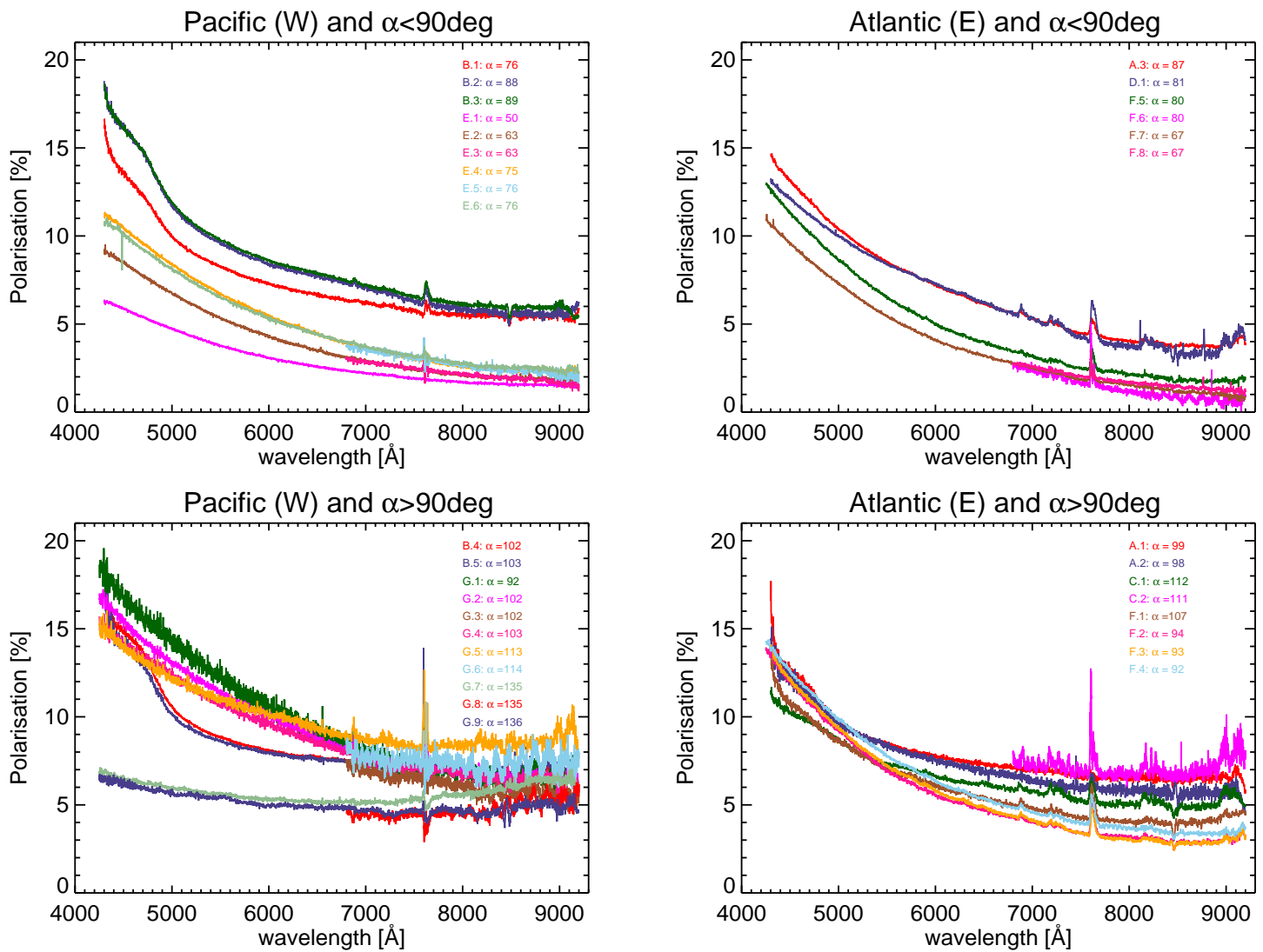} 
\caption{Spectra of the fractional polarization of Earthshine as observed with FORS.  For better visibility, the spectra are grouped as follows. 
         The left two panels show spectra of the Pacific in the Earthshine (to the West of 
         the observing site), and the right two panels of the Atlantic (to the East). 
         The upper panels show spectra taken at phase angles $\alpha < 90$\degr, while the 
         lower panels show those at $\alpha > 90$\degr. Individual observations are identified 
         by colors specified in the legend, and each spectrum identified with its observation ID and its
         actual phase angle in Tab.~\ref{Tab:Log}.}
\label{Fig:AllSpectra}
\end{figure*}

We have extracted polarization spectra for all observations listed in Table \ref{Tab:Log}  consistently applying the methodology described in Sect.~\ref{Sect_Data_Reduction}. 
Figure \ref{Fig:AllSpectra} displays the polarization spectra separated in four groups: viewing sceneries pacific and atlantic, for viewing phase angles 
$\alpha$ above and below 90\degree. Following sect.\ref{DA}, the fractional polarization $P$ is displayed versus wavelength, ranging from 4200 -- 9200~\AA. 
The formal error of the polarization spectra mainly depends on the net signal obtained (dark moon minus background extrapolation). Summing up the signal along the slit direction, and for all retarder settings, we typically deal with total counts above 10$^6$ per wavelength bin, allowing a formal statistical accuracy of below 1\textperthousand \, for all polarization spectra discussed here. 
Spectra obtained with the 600I sometimes exhibit residual "fringing patters" red-ward of 8000~\AA \, possibly caused by variability of telluric sky lines. 
These patterns could not be removed by our flat-fielding procedure. 

Most polarization spectra are smoothly decreasing with increasing wavelength (except for molecular bands, see below). The spectra for datasets with IDs B.x appear to have a broad "bump" in the blue spectral range, around 5000\AA.  We verified that the flatfield procedure did not introduce artefacts, but we cannot exclude some systematic errors for these datasets in this spectral regime.

The different shapes of the individual polarization spectra as shown in 
Fig.~\ref{Fig:AllSpectra} hamper their systematic comparison. But the main shape factors that 
distinguish the spectra can easily be identified: \\
(a) the absolute value of the polarization at a certain wavelength, \\
(b) the slope of the polarization across certain wavelength bands, in particular in the blue 
and the red, \\
(c) the shape and strength of the O$_2$-A band (7600~\AA) and to a lesser extent 
the much weaker O$_2$-B band (6700~\AA). 

Certain simplifications of the description of polarization spectra have been introduced before:  \citet{Bagnulo:2014ey} suggested to "normalize" different polarization spectra at a certain wavelength (in their case at 5500 \AA), in order to allow to compare spectra of different asteroids observed at different phase angles. In their case, such a normalization removes
the phase angle dependence of their observations and hence enables the grouping of the observations 
in different object classes that are otherwise hard to find.

For our purpose, we simplify the description of the detailed spectral shape of individual observations by introducing mean polarization values for different spectral bandpasses.   The mid-points of the spectral bandpasses are derived from the usual astronomical (Johnson) filter system, which are centered around 4450~\AA ($B$), 5510~\AA ($V$), 6580~\AA ($R$), and 8060~\AA ($I$).  A single observation with grism 300V allows us to extract the four mean polarization values and allows an accurate differential measurement of the polarization values in these bands {\sl simultaneously}. In practice, and allowing for inclusion of data from other sources, we have defined 200 to 600~\AA \; wide bandpasses across which we average the measured degree of polarization. The bandpasses chosen are $P_B$: 4350 -- 4550~\AA,  $P_V$: 5450 -- 5650~\AA, $P_R$: 6450 -- 6650~\AA, $P_I$: 8050 -- 8650~\AA. 
The choice of a wide bandpass for $P_I$ is motivated to extend the usual $I$-bandpass to enable future comparison with
data obtained with the POLDER satellite instrument \citep{Deschamps:1994gl}, the only instrument that provides 
polarization measurement from space. POLDER measures polarization only in its reddest channel 
(centered around 865~nm). The choice of  "spectral bands" allows a rather accurate determination 
of the mean polarization value within the band, because the spectral slope across these relatively 
narrow bands is almost constant, and the average value essentially eliminates the residual, 
statistical variations within the band. The errors in the polarization values are thus given by 
their standard deviation measured in their passbands, and are typically very small ($<$ 0.1\textperthousand). 

Table \ref{Tab:Log} lists the values of degree of polarization and their statistical errors for all spectra observed. The spectral coverage of the 600I grism  only allows determination of $P_I$.
For comparison, table \ref{Tab:Log} includes values determined from various sources: low resolution spectra of \citet{2013PASJ...65...38T} 
cover the wavelength range of 4500~\AA \, to 8500 \AA \, and allow to determine the polarization values in all four bandpasses defined above. Their observations consistently contained the African and Asian continents in Earthshine.
The values reported in \citet{2013A&A...556A.117B} refer to measurements of polarization in specific bandpasses, focussing on two regions on the Moon having distinct albedos (lunar highlands and mare). They use the standard Bessell $B, V, R, I$ filter-set. Unlike our definition of $P_I$ that spans a region from 8050-8650 \AA, their $I$ is centered around 8000 \AA.  We estimate the value of polarization $P_I$, assuming linear extrapolation from their $V, R$ and $I$ colors to a wavelength of 8550 \AA. This should reduce bias in this band to compare with our data.

Within the same bandpasses defined above, we can calculate the angle of polarization $\phi_B, \phi_V, \phi_R$ and $\phi_I$. For a Rayleigh scattering atmosphere, this angle should coincide with the normal of the scattering plane on Earth, which is defined by the Sun, observer (on Earth) and the Moon, with a reference direction of the north celestial pole. We have calculated this angle $\Phi$ using Eq.~5 from \citet{Bagnulo:2006ep} for each geometrical configuration of the Sun-Earth-Moon system, and list it in Table \ref{Tab:Log}. As expected, the values of $\phi_B, \phi_V, \phi_R$ and $\phi_I$ have at most only a weak dependance on the wavelength, and coincide quite well with $\Phi$ as calculated. A similar behavior has been found in \citet{2013PASJ...65...38T} and serves as an important sanity check of the quantities $P_Q(\lambda)$ and $P_U(\lambda)$ measured.

\begin{longtab}
\tiny
\begin{landscape}
\begin{longtable}{llrlrrrrrrrrrrrrr}
\caption{Record of Observations: date and time, airmass, grism, exposure times, S--E--M phase angle  $\alpha$, and aspect/geometry ("P" = pacific side, "A" = atlantic side). $\Phi$ is the angle between the normal of the nominal (geometrical) scattering plane and the celestial north pole. The degree of polarization of Earthshine $P^{ES}$, and the angle of polarization $\phi$ as defined in eqs. \ref{eq:pol} and \ref{eq:ang} is tabulated for the bandpasses $B, V, R$ and $I$. Standard errors of the mean are given in brackets, and are -typically - negligible.}\\
\hline\hline
ID & UT-START & AM & Grism & DIT$\times$N$_{cyc}$ & $\alpha$ & Geo.  & $P_B^{\rm ES}[\%]$ & $P_V^{\rm ES}$[\%] & $P_R^{\rm ES}$[\%] &$ P_I^{\rm ES}$[\%]  &
$\Phi$ & $\phi_B$ & $\phi_V$ &$\phi_R$ &$\phi_I$ \\[1mm]
\hline 
\endfirsthead\caption{continued.}\\
\hline\hline 
ID & UT-START & AM & Grism & DIT$\times$N$_{cyc}$ & $\alpha$ & Geo.  & $P_B^{\rm ES}[\%]$ & $P_V^{\rm ES}$[\%] & $P_R^{\rm ES}$[\%] &$ P_I^{\rm ES}$[\%]  &
$\Phi$ & $\phi_B$ & $\phi_V$ &$\phi_R$ &$\phi_I$ \\[1mm]
\hline 
\endhead
\hline
\endfoot
A.1 & 2011-04-24T07:40 & 1.3 & 300V &   20$\times$16 &  99 & A & 12.9(0.04)  &  8.4(0.01)  &  7.3(0.01)  &  6.5(0.01)  &  79 &  79.4(0.05)  &  78.5(0.02)  &  77.9(0.03)  &  76.6(0.04)  \\
A.2 & 2011-04-24T10:10 & 1.0 & 300V &   60$\times$16 &  98 & A & 12.6(0.04)  &  8.2(0.01)  &  6.9(0.01)  &  5.7(0.01)  &  79 &  79.9(0.04)  &  79.1(0.02)  &  78.0(0.05)  &  76.4(0.06)   \\
A.3 & 2011-04-25T08:49 & 1.2 & 300V &  150$\times$16 &  87 & A & 13.4(0.03)  &  8.4(0.02)  &  6.1(0.01)  &  3.9(0.01)  &  75 &  75.6(0.02)  &  75.0(0.01)  &  74.9(0.01)  &  74.3(0.02)   \\[1mm]
B.1 & 2011-06-08T00:32 & 1.6 & 300V &  100$\times$16 &  76 & P & 14.0(0.04)  &  8.2(0.01)  &  6.6(0.01)  &  5.5(0.01)  & 111 & 111.7(0.02)  & 111.5(0.01)  & 110.9(0.02)  & 109.4(0.03)   \\
B.2 & 2011-06-08T23:02 & 1.1 & 300V &   90$\times$16 &  88 & P & 16.5(0.04)  &  9.5(0.02)  &  7.6(0.01)  &  5.6(0.01)  & 112 & 112.2(0.02)  & 112.1(0.01)  & 111.6(0.02)  & 110.5(0.03)   \\
B.3 & 2011-06-09T00:30 & 1.3 & 300V &  180$\times$16 &  89 & P & 16.5(0.04)  &  9.7(0.01)  &  7.8(0.01)  &  6.0(0.01)  & 112 & 112.3(0.02)  & 112.1(0.01)  & 111.6(0.02)  & 110.9(0.03)   \\
B.4 & 2011-06-10T00:47 & 1.1 & 300V &  180$\times$16 & 102 & P & 15.2(0.04)  &  8.8(0.01)  &  7.7(0.01)  &  6.8(0.01)  & 112 & 111.3(0.02)  & 111.2(0.01)  & 110.3(0.02)  & 109.3(0.04)   \\
B.5 & 2011-06-10T01:55 & 1.3 & 300V &  300$\times$16 & 103 & P & 14.4(0.05)  &  8.5(0.01)  &  7.6(0.01)  &  6.9(0.01)  & 112 & 110.9(0.04)  & 111.0(0.02)  & 110.2(0.03)  & 108.8(0.05)  \\[1mm]
C.1 & 2012-10-06T07:12 & 1.7 & 300V &  150$\times$16 & 112 & A & 10.5(0.02)  &  7.3(0.01)  &  6.1(0.01)  &  5.0(0.02)  &  87 &  87.3(0.03)  &  87.1(0.02)  &  87.0(0.03)  &  87.6(0.04)   \\
C.2 & 2012-10-06T08:26 & 1.5 & 600I &  210$\times$16 & 111 & A &  -  &  -  &-  &  6.9(0.00)  &  87 &   -  &   -  &  -  &  87.6(0.00)   \\[1mm]
D.1 & 2012-12-07T08:11 & 1.5 & 300V &  150$\times$16 &  81 & A & 12.4(0.03)  &  8.3(0.01)  &  6.1(0.01)  &  3.6(0.02)  & 112 & 113.0(0.01)  & 112.3(0.01)  & 111.1(0.03)  & 104.6(0.08)   \\[1mm]
E.1 & 2012-12-17T00:19 & 2.1 & 300V &   60$\times$16 &  50 & P &  6.0(0.01)  &  3.7(0.01)  &  2.5(0.01)  &  1.6(0.00)  &  68 &  67.5(0.02)  &  66.3(0.02)  &  64.8(0.03)  &  61.1(0.03)   \\
E.2 & 2012-12-18T00:04 & 1.5 & 300V &   60$\times$16 &  63 & P &  8.7(0.02)  &  5.2(0.01)  &  3.4(0.01)  &  2.0(0.00)  &  66 &  66.1(0.02)  &  65.2(0.02)  &  64.1(0.03)  &  60.8(0.05)   \\
E.3 & 2012-12-18T00:49 & 2.0 & 600I &   90$\times$16 &  63 & P &  -  &  -  & -  &  2.0(0.00)  &  66 &   -  &   -  &  -  &  60.0(0.00)   \\
E.4 & 2012-12-19T00:15 & 1.4 & 300V &   60$\times$16 &  75 & P & 10.7(0.02)  &  6.6(0.02)  &  4.3(0.01)  &  2.4(0.00)  &  66 &  65.7(0.02)  &  65.0(0.01)  &  64.1(0.03)  &  61.1(0.04)   \\
E.5 & 2012-12-19T01:01 & 1.7 & 600I &   90$\times$12+60$\times$4  &  76 & P &  - &  -  & -  &  2.4(0.00)  &  66 &  -  &   -  &  -  &  60.4(0.00)   \\
E.6 & 2012-12-19T01:43 & 2.2 & 300V &   60$\times$12+30$\times$4 &  76 & P & 10.3(0.04)  &  6.4(0.01)  &  4.3(0.01)  &  2.6(0.00)  &  66 &  65.4(0.10)  &  64.9(0.02)  &  63.8(0.03)  &  61.0(0.05)   \\[1mm]
F.1 & 2013-02-02T05:44 & 1.7 & 300V &  120$\times$16 & 107 & A & 11.1(0.04)  &  6.9(0.01)  &  5.3(0.01)  &  4.0(0.01)  & 111 & 112.3(0.03)  & 111.5(0.03)  & 111.4(0.04)  & 112.3(0.07)   \\
F.2 & 2013-02-03T06:39 & 1.6 & 300V &  120$\times$16 &  94 & A & 12.4(0.04)  &  6.8(0.02)  &  4.6(0.01)  &  3.0(0.01)  & 108 & 108.7(0.02)  & 108.7(0.03)  & 109.3(0.04)  & 111.3(0.05)   \\
F.3 & 2013-02-03T07:37 & 1.3 & 300V &  120$\times$16 &  93 & A & 12.5(0.04)  &  7.1(0.02)  &  4.8(0.01)  &  2.9(0.01)  & 108 & 108.7(0.02)  & 108.8(0.02)  & 109.4(0.03)  & 111.7(0.05)   \\
F.4 & 2013-02-03T08:37 & 1.1 & 300V &  120$\times$16 &  92 & A & 12.9(0.03)  &  7.6(0.02)  &  5.2(0.01)  &  3.5(0.01)  & 108 & 108.5(0.02)  & 108.6(0.02)  & 109.4(0.02)  & 111.4(0.04)   \\
F.5 & 2013-02-04T07:17 & 1.8 & 300V &  120$\times$16 &  80 & A & 11.6(0.03)  &  6.3(0.02)  &  3.9(0.01)  &  1.9(0.01)  & 103 & 104.0(0.01)  & 103.2(0.02)  & 102.7(0.04)  & 102.1(0.07)   \\
F.6 & 2013-02-04T08:14 & 1.4 & 600I &  180$\times$16 &  80 & A &  -  & -  & -  &  0.9(0.00)  & 103 &  -  &   -  &  -  & 114.9(0.00)   \\
F.7 & 2013-02-05T07:23 & 2.5 & 300V &  120$\times$16 &  67 & A &  9.8(0.03)  &  5.3(0.02)  &  3.1(0.01)  &  1.3(0.01)  &  99 &  99.8(0.01)  & 100.2(0.02)  & 101.3(0.03)  & 106.4(0.08)   \\
F.8 & 2013-02-05T08:20 & 1.7 & 600I &  180$\times$16 &  67 & A &  -  &  -  & -  &  1.5(0.00)  &  99 &  -  &   -  &  -  & 103.1(0.00)   \\[1mm]
G.1 & 2013-02-18T02:03 & 2.4 & 300V &   60$\times$16 &  92 & P & 17.1(0.05)  & 12.1(0.04)  &  9.0(0.03)  &  6.5(0.02)  &  78 &  79.1(0.06)  &  78.2(0.06)  &  77.3(0.09)  &  75.9(0.09)   \\
G.2 & 2013-02-19T00:06 & 1.4 & 300V &  120$\times$16 & 102 & P & 15.7(0.03)  & 11.2(0.02)  &  8.6(0.01)  &  6.7(0.01)  &  82 &  81.0(0.03)  &  79.1(0.03)  &  77.4(0.04)  &  74.7(0.04)   \\
G.3 & 2013-02-19T01:02 & 1.6 & 600I &  180$\times$16 & 102 & P &  -  & -  & -  &  5.8(0.00)  &  82 &   -  &   -  &  -  &  74.2(0.00)   \\
G.4 & 2013-02-19T02:16 & 2.0 & 300V &  120$\times$16 & 103 & P & 14.4(0.03)  & 10.7(0.02)  &  8.4(0.01)  &  6.5(0.01)  &  82 &  82.1(0.04)  &  80.3(0.03)  &  79.0(0.04)  &  77.3(0.05)   \\
G.5 & 2013-02-20T00:26 & 1.4 & 300V &  120$\times$16 & 113 & P & 14.2(0.03)  & 10.9(0.02)  &  9.3(0.02)  &  8.5(0.02)  &  86 &  84.1(0.04)  &  81.2(0.03)  &  79.1(0.04)  &  76.1(0.05)   \\
G.6 & 2013-02-20T01:38 & 1.5 & 600I &  180$\times$16 & 114 & P &  -  &  -  & -  &  7.2(0.00)  &  87 &   -  &   -  &  -  &  78.3(0.00)   \\
G.7 & 2013-02-22T01:11 & 1.4 & 300V &   90$\times$16 & 135 & P &  6.6(0.01)  &  5.6(0.01)  &  5.1(0.01)  &  5.9(0.01)  &  94 &  91.3(0.04)  &  88.4(0.03)  &  85.9(0.04)  &  84.9(0.04)   \\
G.8 & 2013-02-22T01:58 & 1.4 & 600I &  180$\times$5+150$\times$11 & 135 & P &  -  &  -  & -  &  4.8(0.00)  &  94 &   -  &   -  &  -  &  84.5(0.00)   \\
G.9 & 2013-02-22T03:03 & 1.4 & 300V &   80$\times$16 & 136 & P &  6.3(0.02)  &  5.4(0.01)  &  4.8(0.01)  &  4.6(0.02)  &  94 &  92.6(0.04)  &  90.2(0.04)  &  89.3(0.06)  &  87.3(0.08)   \\[1mm]
\hline 
T1\footnote{from \citet{2013PASJ...65...38T}. Their Earthshine observations contained mainly the continents Asia and Africa, and is denoted by "AA".}  & 2011-03-09T09:55 &  &  &  & 49 &  AA  & 5.9	& 4.1	& 2.9	& 2.1	& 157 & 154(5)\footnote{average (error)  over entire spectrum from 4500 - 8500 \AA }&  &  & 	\\
T2    & 2011-03-10T10:09 &  &  &  & 60 &  AA  &8.8		& 6.3		& 4.3		& 2.7		& 162 & 160(2) &  &  & 	\\
T3    & 2011-03-11T10:18 &  &  &  & 72 &  AA  &10.2	  	& 7.3		& 5.3		& 3.8		& 168 & 167(2) &  &  & 	\\
T4    & 2011-03-12T11:03 &  &  &  & 84 &  AA  &9.8		& 7.6		& 6.0		& 4.8		& 174 & 171(1) &  &  & 	\\
T5    & 2011-03-13T10:27 &  &  &  & 96 &  AA  &9.2		& 7.5		& 6.1		& 5.3		& 179 & 177(1) &  &  & 	\\[1mm]
\hline 
B1H\footnote{from \citet{2013A&A...556A.117B}. "H" means that Earthshine was reflected from lunar Highlands while "M" from lunar Mare.} & 2011-10-02T &   &  &  & 73 &  H  & 8.7		& 6.1		& 4.0		&  2.5\footnote{$P_I$ values extrapolated to a wavelength of 8550 \AA } & n.a. &  &  &  &   \\
B1M & 2011-10-02T &  &  &  & 73 &  M  & 11.9		& 8.1		& 5.3		&  2.9$^{3}$ & n.a. &  &  &  & \\
B2H & 2011-10-03T &  &  &  & 73 &  H  & 9.4		& 6.9		& 4.8		&  2.9$^{3}$ & n.a. &  &  &  & \\
B2M & 2011-10-03T&   &  &  & 73 &  M & 12.7		& 8.6		& 5.7		&  2.7$^{3}$ & n.a. &  &  &  & \\
B3H & 2011-10-04T &  &  &  & 73 &  H  & 9.1		& 6.9		& 5.0		&  3.4$^{3}$ & n.a. &  &  &  & \\
B3M & 2011-10-04T &  &  &  & 73 &  M  & 11.9		& 8.6		& 6.1		&  3.7$^{3}$ & n.a. &  &  &  & \\
B4H & 2011-10-05T &   &  &  & 73 & H  & 7.7		& 5.7		& 4.2		&  3.1$^{3}$ & n.a. &  &  &  & \\
B4M & 2011-10-05T &  &  &  & 73 &  M  & 11.1		& 8.4		& 5.9		&  8.6$^{3}$ & n.a. &  &  &  &  \\[1mm]
\hline
\hline
\label{Tab:Log} 
\end{longtable}
\end{landscape}
\end{longtab}

\begin{figure} 
\resizebox{\hsize}{!}{\includegraphics{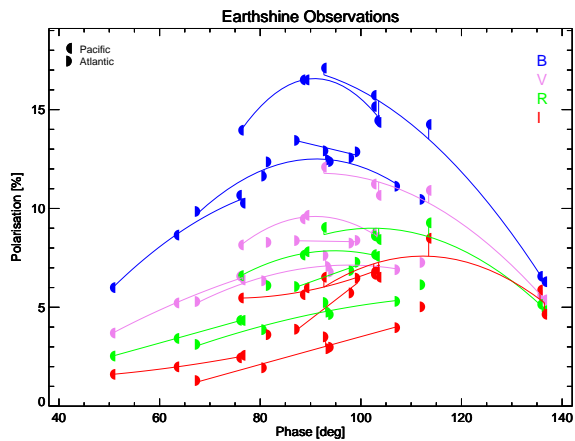}} 
\caption{Measured fractional polarization of Earthshine as a function of the phase angle 
         (Sun -- Earth -- Moon). The different symbols refer to different viewing geometries 
         during the observations: the Pacific or Atlantic ocean. The bandpasses used correspond 
         approximately to the traditional $B$, $V$, $R$ and $I$ bandpasses (see text). The lines 
         connect at least three independent observation cycles that belong to a consecutive 
         sequence of observations within a given observing run (see Tab.~\ref{Tab:Log}).}
\label{Fig:AllPA_own}
\end{figure}

\subsection{Polarization Phase Curves}\label{Phasecurve}

Astronomical objects (with or without any atmosphere) exhibit a characteristic variation of their polarization with phase angle  \citep[see e.g.][]{2015psps.book.....K}. 
Already since \citet{1957SAnAp...4....3D} it is known that the observed polarization of Earthshine follows a characteristic phase curve. 
Figure \ref{Fig:AllPA_own} shows our measured polarization values at phase angles
from 50\degree\ to 140\degree\ (see Tab.~\ref{Tab:Log}). The four passbands ($B$, $V$, $R$ and $I$) have
been indicated by, respectively, blue, violet, green and red lines. The two different types of 
symbols distinguish the viewing geometry of the Earth during the observations: left-half circles
are observations of the Atlantic (east to the observational site in Chile), while right-half circles
are of the Pacific (west to the observational site). This distinction is important as it indicates
both the different global sceneries, and the fact that a different lunar limb region was observed
at the time. One set of polarization measurements in the four bands at a given phase angle and
with a given viewing geometry (Pacific or Atlantic) represents a single Earthshine polarization spectrum. 

In order to simplify the interpretation of the cluster of points in Fig.~\ref{Fig:AllPA_own}, 
we have connected at least three independent observation cycles that belong to a continuous sequence of observations in a given band and observing epoch, by lines using {\bf a second order polynomial for} a
least-square fit procedure.   The lines highlight the overall shape and grouping of polarization 
values. Individual observations that belong to one sequence of observations within one observing 
run lie in general close to the connecting line. The polarization generally
follows a smooth curve as a function of the phase angle. The polarization reaches its highest 
value roughly around $\alpha= 80$\degree\ to 100\degree\ (i.e.\ around quadrature) and decreases towards 
lower and higher phase angles within our observational range. There is a tendency for longer wavelengths 
to have the polarization maximum at larger phases angles.

There is a significant dispersion of the polarization when comparing distinct observation epochs, 
in all bands. The variation of the polarization values appears to be largest around quadrature. 
For example, observations of the Pacific side around quadrature exhibit maximum polarization 
values in the blue of more than 16\% (B.2, B.3 and G.1), while observations at a similar geometry, 
of the Atlantic side are about 3\% lower (F.4).  
In general, the polarization appears to be lower for observations on the Atlantic side than
on the Pacific side. This was already noted by \citet{Sterzik:2012gk} for datasets A.3 and B.4, 
and interpreted in terms of different cloud fractions at the time of the observations, with higher polarization corresponding to a lower cloud coverage fraction. 
A further important difference between the Pacific and the Atlantic observations is the highly polarized sunglint on the ocean, 
which is mostly visible on the Pacific side and only partly on the Atlantic side \citep{Emde:2017ee}.

\subsubsection{Comparison with other observations of Earthshine polarization}

\begin{figure} 
\resizebox{\hsize}{!}{\includegraphics{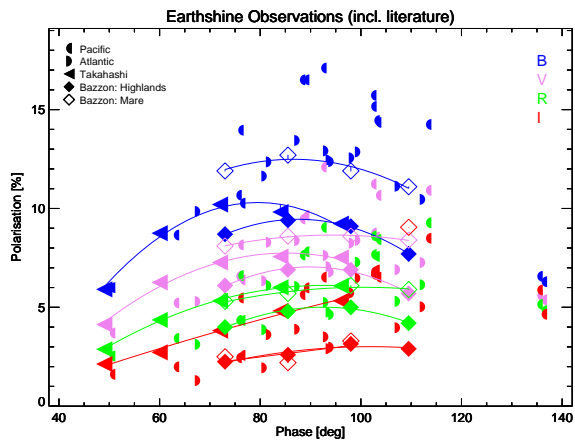}} 
\caption{Polarization of Earthshine as a function of phase angle $\alpha$.
         The different symbols refer to different observations, including observations by 
         \citet{2013PASJ...65...38T, 2013A&A...556A.117B}. The bandpasses correspond approximately
         to $B$, $V$, $R$, and $I$ bandpasses, as described in the text.
}
\label{Fig:AllPA}
\end{figure}

How do the measurements relate to observations published before? For comparison, Figure \ref{Fig:AllPA} shows values from the literature as listed in Tab.~\ref{Tab:Log}. 
Our Earthshine measurement can most directly be compared to those of  \citet{2013PASJ...65...38T} and \citet{2013A&A...556A.117B}.  \citet{2013A&A...556A.117B} report polarization values in certain passbands.  From the spectra of \citet{2013PASJ...65...38T} we derive the passband values with the same procedure as for our own spectra.  Similarly to above, we connect their data by lines following  least-square fits, as all observations fall within a few days. In general, their polarization values  follow the same trends as ours. 
The data that belong to one continuous set of observations show little 
scatter around the fitted lines. However, compared amongst each other, and with our data, 
the scatter in the Earthshine polarization data of different authors obtained at different epochs 
but at similar phase angles, reaches around 2\%.

An important source of scatter of the polarization data was clearly identified in \citet{2013A&A...556A.117B}, who measured Earthshine polarization in two different regions on the lunar surface (on Highlands and in Mare), quasi simultaneously. The different surface types covering these regions (their albedos are 0.21 
and 0.11, respectively) appear to yield significantly different lunar depolarization factors, and hence
a relative difference of the measured polarization of the Earthshine of up to 30\% in the blue ($B$) 
spectral band.

In the next section, we investigate the effect of lunar depolarization on our data.

\subsubsection{Correction for lunar depolarization}\label{depol}

Earthshine that we measure has been reflected by the lunar surface. Depending
on the composition and structure of the local lunar surface, the reflection changes the 
state of polarization of the Earthshine.
The lunar depolarization factor or polarization efficiency $\epsilon$ is defined as 
\begin{equation}
    \epsilon(\lambda) = {P^{\rm out}(\lambda)}/ \; {P^{\rm in}(\lambda)}, 
\label{eq_epsilon}
\end{equation}
with $P^{\rm in}$ the polarization of the light that is incident on the moon, and $P^{\rm out}$ the 
polarization of the reflected light. Note that in order to measure $\epsilon$, one could 
illuminate lunar (analogue) surface samples with light with a known polarization state, and
measure the polarization of the reflected light for a wide range of illumination and viewing 
geometries and wavelengths. As far as we know, such measurements have not been done yet.

\citet{2013A&A...556A.117B} introduce a method to correct Earthshine polarimetry using the knowledge 
of the lunar albedo at the location where the Earthshine is observed, assuming that $\epsilon$
depends on the wavelength and the lunar surface albedo. In addition, it is assumed that
the phase angle dependence of $\epsilon$ is negligible because the angle between the lit part of Earth and the observer, as seen from the moon, is small and always around 1\degree$\pm$0.5\degree. The spectral dependence 
of the polarization efficiencies on albedo were derived through an analysis of lunar samples by  
\citet{1993Sci...260..509H}. Finally, lunar albedos are obtained from an extrapolation of absolute lunar albedos maps by \citet{Velikodsky:2011dg} to backscatter angles of 1\degree.
The polarization efficiency, $\log \epsilon$, as a function of the lunar albedo at 603~nm, $\log a_{603}$, 
and the wavelength, $\log \lambda$, as derived by \citet{2013A&A...556A.117B} is then
\begin{equation} 
\label{eq:depol}
    \log \epsilon (\lambda, a_{603}) = -0.61 \log a_{603} - 0.291 \log \lambda [\mu m] - 0.955. 
\end{equation}

We use the same approach to derive the polarization efficiency for our observations. The acquisition images obtained immediately before the spectropolarimetric measurements were used to identify their location on the lunar surface, and we matched them  with the extrapolated 1-deg albedo maps by  \citet{Velikodsky:2011dg}. This can be seen in Fig.~\ref{Fig:acq} where we have scaled and rotated the albedo map to best match the orientation of the acquisition image. The determination of the rotation angle $\alpha_{\rm Moon}$ with this procedure is affected by field distortion, the finite spatial resolution of the images, and the sometimes low contrast of the acquisition images. We estimate the accuracy of the rotation is limited to $\pm$ 1\degree, while the accuracy to locate and extract the correct albedo values at the position of the slitlets is accurate to no more then $\pm 15-20 \arcsec$ on the moon.  We therefore sampled the albedos  $a_{603}$ scanning the slit mask $\pm 5$ pixels (corresponding to $\pm 18\arcsec$ on the moon) around its suspected center position in and perpendicular to the slit direction. We then derived a mean albedo $\overline{a}_{603}$ for all 11$\times$11 raster slit images, and estimated the error of the albedo measurement using the standard deviation within these samples.   In Tab.~\ref{Tab:PEarth} we list the rotation angle $\alpha_{Moon}$   (counted counterclockwise from the North) and the mean albedo $\overline{a}_{603}$ and its standard deviation derived from these positions for each observation.

\begin{table*}
\caption{Degree of polarization $P^{\rm E}$ in our bandpasses corrected for lunar depolarization. The determination of rotation angle $\alpha_{\rm Moon}$ of the moon 
         with respect to the acquistion images and mean albedo $\overline{a}_{603}$ for 
         each observed region are explained in the text. Systematic errors of $P^{\rm E}$ are derived from the errors of $\overline{a}_{603}$. Also listed with their errors are the 
         equivalent width of the O$_2$-A band and the differential polarization vegetation index 
         $\Delta$PVI.}
\label{Tab:PEarth} 
\centering
\begin{tabular}{lrlrrrrrcl}
\hline\hline
ID & $\alpha_{\rm Moon}$ & $\overline{a}_{603}$ & $P_B^{{\rm E}}$[\%] & $P_V^{{\rm E}}$[\%] & $P_R^{{\rm E}}$[\%] &$ P_I^{{\rm E}}$[\%]  & EW(O$_2$-A)[\AA] & $\Delta$PVI [\textperthousand] \\
\hline 
A.1 & 275.0 & 0.176(0.004) & 31.8$^{+ 0.8}_{- 0.8}$ & 22.1$^{+ 0.6}_{- 0.6}$ & 20.1$^{+ 0.5}_{- 0.5}$ & 19.2$^{+ 0.5}_{- 0.5}$ &   -1.3(0.1) &  -1.4(0.2) \\
A.2 & 273.0 & 0.169(0.003) & 30.3$^{+ 0.7}_{- 0.7}$ & 21.2$^{+ 0.5}_{- 0.5}$ & 18.5$^{+ 0.4}_{- 0.4}$ & 16.6$^{+ 0.4}_{- 0.4}$ &   -4.4(0.2) &  -2.3(0.3) \\
A.3 & 277.5 & 0.173(0.005) & 32.9$^{+ 1.2}_{- 1.2}$ & 21.8$^{+ 0.8}_{- 0.8}$ & 16.6$^{+ 0.6}_{- 0.6}$ & 11.4$^{+ 0.4}_{- 0.4}$ &  -10.5(0.1) &  -1.9(0.1) \\[1mm]
B.1 & 140.0 & 0.176(0.007) & 34.5$^{+ 1.6}_{- 1.7}$ & 21.5$^{+ 1.0}_{- 1.1}$ & 18.3$^{+ 0.9}_{- 0.9}$ & 16.2$^{+ 0.8}_{- 0.8}$ &   -3.5(0.1) &  -1.0(0.1) \\
B.2 & 138.5 & 0.177(0.006) & 40.9$^{+ 1.8}_{- 1.9}$ & 25.0$^{+ 1.1}_{- 1.1}$ & 21.2$^{+ 0.9}_{- 1.0}$ & 16.8$^{+ 0.7}_{- 0.8}$ &   -6.1(0.1) &  -2.4(0.2) \\
B.3 & 138.5 & 0.177(0.006) & 40.9$^{+ 1.7}_{- 1.7}$ & 25.5$^{+ 1.1}_{- 1.1}$ & 21.7$^{+ 0.9}_{- 0.9}$ & 17.8$^{+ 0.7}_{- 0.8}$ &   -6.7(0.0) &  -1.3(0.1) \\
B.4 & 140.5 & 0.179(0.007) & 37.8$^{+ 1.8}_{- 1.8}$ & 23.3$^{+ 1.1}_{- 1.1}$ & 21.4$^{+ 1.0}_{- 1.0}$ & 20.4$^{+ 1.0}_{- 1.0}$ &   -3.7(0.1) &  -3.0(0.2) \\
B.5 & 137.5 & 0.179(0.007) & 36.0$^{+ 1.6}_{- 1.7}$ & 22.7$^{+ 1.0}_{- 1.1}$ & 21.2$^{+ 1.0}_{- 1.0}$ & 20.6$^{+ 0.9}_{- 1.0}$ &   -8.9(0.0) &  -2.2(0.2) \\[1mm]
C.1 & 276.0 & 0.171(0.006) & 25.4$^{+ 1.2}_{- 1.2}$ & 18.8$^{+ 0.9}_{- 0.9}$ & 16.7$^{+ 0.8}_{- 0.8}$ & 14.6$^{+ 0.7}_{- 0.7}$ &  -13.2(0.2) &  -3.2(0.2) \\
C.2 & 276.0 & 0.172(0.007) &  - &  - & - & 20.0$^{+ 0.9}_{- 0.9}$ &  -17.3(0.0) &  -6.8(0.3) \\[1mm]
D.1 & 245.0 & 0.173(0.006) & 30.2$^{+ 1.3}_{- 1.3}$ & 21.6$^{+ 0.9}_{- 1.0}$ & 16.7$^{+ 0.7}_{- 0.7}$ & 10.6$^{+ 0.5}_{- 0.5}$ &  -27.9(0.6) &  -3.6(0.2) \\[1mm]
E.1 & 110.0 & 0.174(0.007) & 14.7$^{+ 0.7}_{- 0.7}$ &  9.7$^{+ 0.4}_{- 0.4}$ &  6.9$^{+ 0.3}_{- 0.3}$ &  4.7$^{+ 0.2}_{- 0.2}$ &   -0.8(0.0) &  -0.1(0.0) \\
E.2 & 113.0 & 0.175(0.007) & 21.3$^{+ 1.0}_{- 1.0}$ & 13.7$^{+ 0.6}_{- 0.6}$ &  9.4$^{+ 0.4}_{- 0.4}$ &  5.9$^{+ 0.3}_{- 0.3}$ &   -3.7(0.1) &  -0.0(0.1) \\
E.3 & 113.5 & 0.175(0.007) &  - &  - & - &  5.8$^{+ 0.3}_{- 0.3}$ &   -1.9(0.1) &  -0.7(0.1) \\
E.4 & 118.0 & 0.179(0.007) & 26.6$^{+ 1.2}_{- 1.3}$ & 17.5$^{+ 0.8}_{- 0.8}$ & 12.1$^{+ 0.6}_{- 0.6}$ &  7.3$^{+ 0.3}_{- 0.3}$ &   -8.2(0.1) &  +0.3(0.1) \\
E.5 & 118.0 & 0.181(0.007) &  - &  - &- &  7.4$^{+ 0.3}_{- 0.4}$ &   -1.9(0.0) &  +4.1(0.1) \\
E.6 & 118.0 & 0.183(0.007) & 25.9$^{+ 1.2}_{- 1.3}$ & 17.2$^{+ 0.8}_{- 0.8}$ & 12.3$^{+ 0.6}_{- 0.6}$ &  7.8$^{+ 0.4}_{- 0.4}$ &   -3.9(0.1) &  +0.6(0.1) \\[1mm]
F.1 & 247.0 & 0.183(0.007) & 28.1$^{+ 1.4}_{- 1.4}$ & 18.6$^{+ 0.9}_{- 0.9}$ & 15.0$^{+ 0.7}_{- 0.8}$ & 12.0$^{+ 0.6}_{- 0.6}$ &  -18.5(0.1) &  -0.9(0.2) \\
F.2 & 250.0 & 0.183(0.008) & 31.2$^{+ 1.5}_{- 1.6}$ & 18.3$^{+ 0.9}_{- 0.9}$ & 13.1$^{+ 0.6}_{- 0.7}$ &  9.0$^{+ 0.4}_{- 0.5}$ &  -24.1(0.1) &  -3.2(0.2) \\
F.3 & 250.0 & 0.182(0.008) & 31.4$^{+ 1.6}_{- 1.6}$ & 19.0$^{+ 1.0}_{- 1.0}$ & 13.4$^{+ 0.7}_{- 0.7}$ &  8.8$^{+ 0.4}_{- 0.5}$ &  -19.5(0.1) &  -2.3(0.1) \\
F.4 & 250.0 & 0.182(0.008) & 32.6$^{+ 1.6}_{- 1.7}$ & 20.5$^{+ 1.0}_{- 1.1}$ & 14.8$^{+ 0.7}_{- 0.8}$ & 10.6$^{+ 0.5}_{- 0.6}$ &  -19.7(0.1) &  -1.3(0.1) \\
F.5 & 251.0 & 0.182(0.008) & 29.3$^{+ 1.5}_{- 1.5}$ & 17.0$^{+ 0.9}_{- 0.9}$ & 10.9$^{+ 0.6}_{- 0.6}$ &  5.9$^{+ 0.3}_{- 0.3}$ &  -20.5(0.1) &  -2.2(0.1) \\
F.6 & 251.0 & 0.182(0.008) &  - &  - & - &  2.8$^{+ 0.1}_{- 0.2}$ &  -39.8(0.2) &  +4.2(0.2) \\
F.7 & 258.0 & 0.181(0.008) & 24.7$^{+ 1.3}_{- 1.4}$ & 14.2$^{+ 0.8}_{- 0.8}$ &  8.8$^{+ 0.5}_{- 0.5}$ &  3.9$^{+ 0.2}_{- 0.2}$ &  -16.8(0.2) &  -2.2(0.1) \\
F.8 & 258.0 & 0.181(0.008) &  - &  - & - &  4.5$^{+ 0.2}_{- 0.3}$ &   -9.0(0.3) &  -1.9(0.1) \\[1mm]
G.1 & 103.0 & 0.179(0.010) & 42.7$^{+ 2.7}_{- 2.8}$ & 32.2$^{+ 2.1}_{- 2.1}$ & 25.3$^{+ 1.6}_{- 1.7}$ & 19.6$^{+ 1.2}_{- 1.3}$ &   -5.9(0.2) &  -0.8(0.5) \\
G.2 &  99.0 & 0.179(0.010) & 39.2$^{+ 2.7}_{- 2.8}$ & 29.8$^{+ 2.0}_{- 2.1}$ & 24.0$^{+ 1.6}_{- 1.7}$ & 19.9$^{+ 1.4}_{- 1.4}$ &  -10.3(0.1) &  +0.7(0.2) \\
G.3 &  98.5 & 0.178(0.011) &  - &  - & - & 17.2$^{+ 1.3}_{- 1.3}$ &   -3.0(0.1) &  +1.1(0.2) \\
G.4 &  98.0 & 0.177(0.012) & 35.5$^{+ 2.8}_{- 2.9}$ & 28.2$^{+ 2.2}_{- 2.3}$ & 23.4$^{+ 1.8}_{- 1.9}$ & 19.4$^{+ 1.5}_{- 1.6}$ &   -5.1(0.2) &  +0.9(0.3) \\
G.5 &  93.0 & 0.176(0.012) & 35.2$^{+ 2.8}_{- 2.9}$ & 28.7$^{+ 2.3}_{- 2.4}$ & 25.6$^{+ 2.0}_{- 2.1}$ & 25.2$^{+ 2.0}_{- 2.1}$ &   -9.9(0.1) &  -2.0(0.3) \\
G.6 &  92.0 & 0.175(0.012) &  - &  - & - & 21.2$^{+ 1.7}_{- 1.8}$ &   +3.5(0.2) &  +8.9(0.4) \\
G.7 &  90.0 & 0.174(0.012) & 16.1$^{+ 1.4}_{- 1.4}$ & 14.6$^{+ 1.2}_{- 1.3}$ & 14.1$^{+ 1.2}_{- 1.3}$ & 17.3$^{+ 1.5}_{- 1.5}$ &   +4.1(0.1) &  -0.1(0.2) \\
G.8 &  90.0 & 0.174(0.013) &  - &  - & - & 14.0$^{+ 1.2}_{- 1.3}$ &  +10.5(0.2) &  -0.6(0.2) \\
G.9 &  90.0 & 0.173(0.013) & 15.4$^{+ 1.4}_{- 1.5}$ & 14.0$^{+ 1.3}_{- 1.4}$ & 13.2$^{+ 1.2}_{- 1.3}$ & 13.6$^{+ 1.2}_{- 1.3}$ &   +3.9(0.1) &  -0.7(0.2) \\[1mm]
\hline
E-A.3\tablefootmark{a} &		&		& 29.7	& 18.9	& 13.8	& 8.6		& -13.57(0.04)		& -4.49(0.37)\\
E-B.4 &		&		& 39.7	& 29.1	& 23.3	& 16.7	& +0.94(0.07)		& -1.95(0.23) \\[1mm]
\hline
\hline
\end{tabular}
\tablefoot{
\tablefoottext{a}{simulation values  from \citet{Emde:2017ee}}}
\end{table*}

\begin{figure} 
\resizebox{\hsize}{!}{\includegraphics{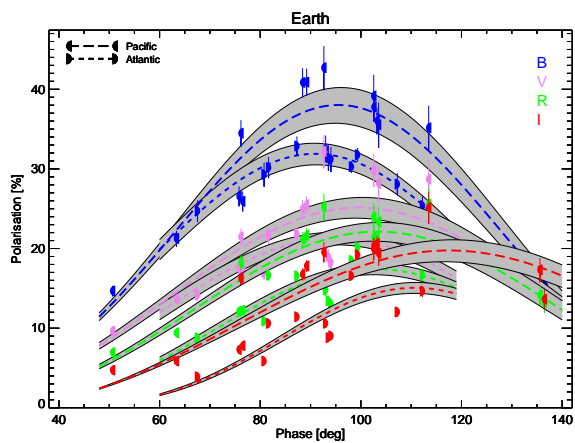}} 
\caption{Polarization of Earth after correction for lunar depolarization effects.  }
\label{Fig:Earthpol}
\end{figure}

We have plotted the lunar depolarization corrected polarizations for all four passbands in 
Fig.~\ref{Fig:Earthpol}. As in Fig.~\ref{Fig:AllPA_own}, different colors indicate different bandpasses, 
and the two different symbols distinguish the observations of the Pacific (P) and the Atlantic (A) 
side of the Earth. The errors derived from applying minimum and maximum polarization efficiencies due to
the scatter in the lunar albedos across the region where the Earthshine is measured are 
shown with vertical bars. 
For each passband and orientation of the Earth ("P" or "A"), the data appears to be scattered around 
a mean phase curve. To fit these phase curves, we use a modified Rayleigh function of the form 
\citep{2005SoSyR..39...45K}
\begin{equation}\label{eq:rayleigh}
P(\alpha) = \frac{(\sin^2 (\alpha - \Delta\alpha))^W}{1 + \cos^2 (\alpha - \Delta\alpha) + dePol}
\end{equation}
Parameters  $W$, $\Delta\alpha$ and $dePol$ characterize each curve with respect to its width and phase shift with respect to $\alpha = 90$\degree. Parameter $dePol$ relates to the maximum polarization. In Fig.~\ref{Fig:Earthpol}, we show the best fits with the 1-$\sigma$ 
confidence intervals indicated with grey bands.
The fit parameters are listed in Tab.~\ref{Tab:Phasefits}. The upper and lower errors are obtained 
by fitting the data with the upper and lower polarization values obtained by assuming different 
polarization efficiencies $\epsilon$ due to systematic albedo variations as listed in 
Tab.~\ref{Tab:PEarth}. 
The datasets over the Pacific and the Atlantic are distinguishable by different fit parameters 
that are statistically significant at the 1--3 $\sigma$ level. It is interesting to note that 
the differences between the sets are more significant in the redder ($R$ and $I$) spectral passbands. 
 Another interesting feature in the curves is the apparent crossing of the polarization curves for the different bands near 120\degree: planet Earth becomes 'white' in polarization. 

\begin{table}
\caption{Values of the best fit parameters for  $W$, $\Delta\alpha$ and $dePol$  (Eq.~\ref{eq:rayleigh})
         for the Pacific and Atlantic datasets in all four passbands. Errors have been propagated 
         from errors in the lunar albedo determination. }
\label{Tab:Phasefits} 
\centering
\begin{tabular}{lrcc}
\hline\hline
 & $\Delta\alpha$ [$\deg$] & $W$ & $dePol$  \\ 
\hline
   \multicolumn{4}{c}{Pacific}\\
   \hline
$B$ &  5.53$^{+0.50}_{-0.61}$ &  1.29$^{+0.03}_{-0.03}$ &  1.63$^{+0.14}_{-0.17}$\\[1mm]
$V$ &  9.73$^{+0.66}_{-0.78}$ &  1.07$^{+0.04}_{-0.05}$ &  2.98$^{+0.19}_{-0.22}$\\[1mm]
$R$ & 13.33$^{+0.73}_{-0.85}$ &  1.16$^{+0.05}_{-0.06}$ &  3.52$^{+0.22}_{-0.26}$\\[1mm]
$I$ & 27.51$^{+0.78}_{-1.02}$ &  0.93$^{+0.02}_{-0.03}$ &  4.06$^{+0.32}_{-0.39}$\\[1mm]
\hline
 \multicolumn{4}{c}{Atlantic}\\
   \hline
$B$ &  0.89$^{+0.20}_{-0.21}$ &  1.28$^{+0.06}_{-0.06}$ &  2.14$^{+0.13}_{-0.14}$\\[1mm]
$V$ &  4.95$^{+0.06}_{-0.07}$ &  1.27$^{+0.10}_{-0.11}$ &  3.70$^{+0.17}_{-0.19}$\\[1mm]
$R$ & 12.78$^{+0.19}_{-0.19}$ &  1.62$^{+0.07}_{-0.08}$ &  4.74$^{+0.22}_{-0.25}$\\[1mm]
$I$ & 20.73$^{+0.18}_{-0.17}$ &  2.33$^{+0.04}_{-0.05}$ &  5.64$^{+0.28}_{-0.32}$\\[1mm]
\hline
\hline
\end{tabular}
\end{table}

The scatter of the observed polarization that is visible in Fig.~\ref{Fig:Earthpol} is caused 
both by intrinsic scatter due to the relative uncertainties of the lunar albedos and the 
associated uncertainty in the lunar polarization efficiency, and by the sensitivity of 
the polarization to the atmospheric and surface properties of the Earth as seen from the Moon
at the time of the observations. 
In particular datapoints outside the error regime of lunar depolarization are likely 
caused by differences in the atmospheric and surface patterns, as they influence the Earth's 
polarization significantly. A comparison with dedicated models should help to understand the 
causes of these deviations.

\subsubsection{Comparison with models}

In this section we compare our measurements with idealized models of polarization spectra of Earth-like planets by \citet{Stam:2008ij}. The results of these numerical models are very useful for a qualitative comparison with our data. They are available in tabulated form with sufficient spectral resolution, for the full phase curve, and for a variety of parameters that characterize the surface and atmospheric properties such as the fraction of free ocean and land surfaces (including vegetated areas), as well as cloud cover.  The tabulated data are limited in
that they apply to horizontally homogeneous model planets, that the reflection by land
surfaces is only Lambertian (i.e. isotropic and depolarizing) and while the reflection by ocean 
surfaces is described by Fresnel reflection, thus including polarization, there are no waves
on the ocean. The glint due to the reflection of the direct, i.e. non-scattered, sunlight 
on the water is thus described by a delta-function, not by an extended region, as it would 
be on a wind--ruffled surface \citep{Zugger:2010dx, Zugger:2011it}. 
The tabulated data further include a single type of 
(liquid water) cloud only, with an optical thickness of ten.
Despite the limitations, these tabulated data can serve as a first approximation to explain 
the most prominent observational features of the Earthshine, and may allow to identify the 
main physical mechanisms that impact the Earthshine spectra, and have been used already 
in \citet{Sterzik:2012gk}. By taking weighted sums of the tabulated data, we mimic
horizontally inhomogeneous planets \citep{Stam:2008ij}.

\begin{figure} 
\resizebox{\hsize}{!}{\includegraphics{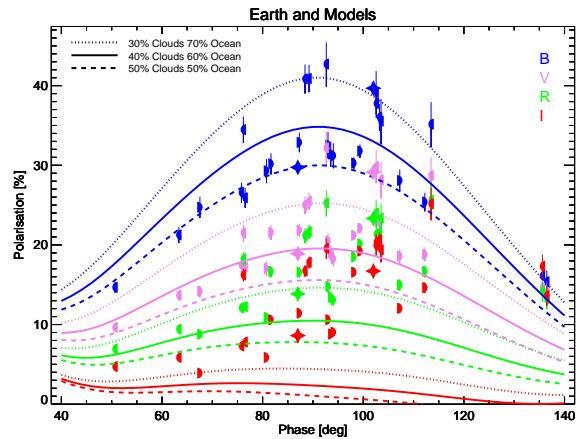}} 
\caption{Comparison of the fractional polarization of Earth in dependency of phase angle (Sun -- Earth -- Moon) with models in B, V, R and I bandpasses. Error bars indicate the effects of uncertain lunar albedo on polarization. Phase curves for three representative models of \citet{Stam:2008ij} are over-plotted with different linestyles.  Values extracted from the simulated spectra of \citet{Emde:2017ee} for observation IDs A.3 and B.4 are indicated  by stars.}
\label{Fig:AllPA_swarm}
\end{figure}

In Fig.~\ref{Fig:AllPA_swarm} we compare Earth's polarization values (that have been individually corrected by polarization efficiencies as described above)  with three representative models of \citet{Stam:2008ij}. 
We consider cloud coverage fractions of 30\%, 40\%, and 50\%, with the remainder of the disk
covered by cloud-free ocean. As expected, in every bandpass, the polarization decreases with 
increasing cloud coverage, because at most phase angles, the polarization due to the clouds
is lower than that due to the atmospheric gas above the dark ocean. The polarization decreases
with increasing wavelength because the scattering by the gas decreases with increasing wavelength. 
The scattering by the clouds thus becomes more prominent with increasing wavelength.
In the red bandpass, $\pq$ changes sign at phase angles larger than 90$^\circ$ with 40\% and 50\% cloud fraction
\footnote{We note that polarization in the models of \citet{Stam:2008ij} is only defined by $\pq$  in the reference plane for which $\pu=0$ which explains negative values in the model curves. }, 
because
at those phase angles, the polarized flux reflected by the clouds has a direction parallel
to the reference plane and the angle of polarization $\phi$ defined in Eq.~\ref{eq:ang} changes by 90\degree\ 
\citep[see][for models at a large range of cloud coverages and
phase angles]{Karalidi:2012fc}. Note that our observations do not include the phase angle range
where the rainbow due to the scattering of light in the spherical water cloud droplets 
is expected: this peak in polarization would appear around $\alpha=40^\circ$, with the 
increase in polarization towards the peak starting below $\alpha=50^\circ$
\citep{Karalidi:2012fc}, just the smallest phase angle in our observations.
The lines for the lower cloud fraction and for the higher cloud fraction bracket the observations 
for the $B$-band and for most observations in the $V$-band. 
The lines for the $R$- and in particular 
the $I$-bands, however, are too low to bracket the observations. 
We conclude that models of the Earth with a pure ocean surface and 
liquid water clouds
bracket the observed polarization phase curves in the blue spectral region. 
Differences in the observed cloud coverage fractions and in particular cloud types (optical thickness, cloud
particle size and possibly thermodynamic phase) for the different observing epochs likely 
cause the different polarization fractions observed during the course of one observation 
epoch covering at most a few days. The red $R$ and in particular $I$ spectral bands remain a challenge to fit
with a weighted sum of horizontally homogeneous cloudy and ocean-covered Earth polarization models.  

The relatively flat continuum of the polarization spectra for red wavelengths as compared to \citet{Stam:2008ij} had already been noted in \citet{Sterzik:2012gk}. This triggered to develop and apply more sophisticated models, in particular for the treatment of clouds. A dedicated model was explained and presented in detail in \citet{Emde:2017ee}.  Their simulation was actually designed to explain observations ID A.3 and ID B.4. 
In order to compare the simulation result with measurements,  we derived polarization values directly from their spectra, and listed them in Table~\ref{Tab:PEarth}. As can be seen in Fig.~\ref{Fig:AllPA_swarm}, their polarization values tend to fit the observations in all four  $B$, $V$, $R$ and $I$ bands.

\subsection{Polarization Color Ratios}\label{ColorRatios}

The comparison of Earthshine polarization with models of Earth polarization is hampered by the uncertainties induced by the rather inaccurate knowledge of the lunar (de-)polarization factor. The method of \citet{2013A&A...556A.117B} applied in Sect.~\ref{depol} is only an approximation. The rather uncertain determination of the exact slit position on the Moon introduces an additional error budget  that we estimated above. But in particular the rather unknown scattering parameters of (different) lunar soil across the Moon may introduce even more systematic errors, and Eq.~\ref{eq:depol} holds only approximately. The absolute polarization derived for Earth from Earthshine may thus be uncertain by a few percent, with relative errors as large as 10\%.

However, Eq.~\ref{eq:depol} suggests only a rather weak dependence of the lunar polarization 
efficiency $\epsilon$ on $\lambda$.  A factor ten difference in wavelength results in a factor of less than two difference in $\epsilon$,  and the relative error introduced when considering polarization ratios such as $P_B/P_V$ (or $P_R/P_I$) is expected to be only of the order of 4-5\%. 
Thus, even 
if Eq.~\ref{eq:depol} may just be an approximation, polarization ratios, in particular pertaining
to adjacent wavelength bands, should be rather robust quantities to compare with models. 
Such differential quantities could thus serve as generic observables of Earthshine that allow 
a more reliable comparison with models, because they are less sensitive to the lunar depolarization
factor and its uncertainties.

\begin{figure} 
\resizebox{\hsize}{!}{\includegraphics{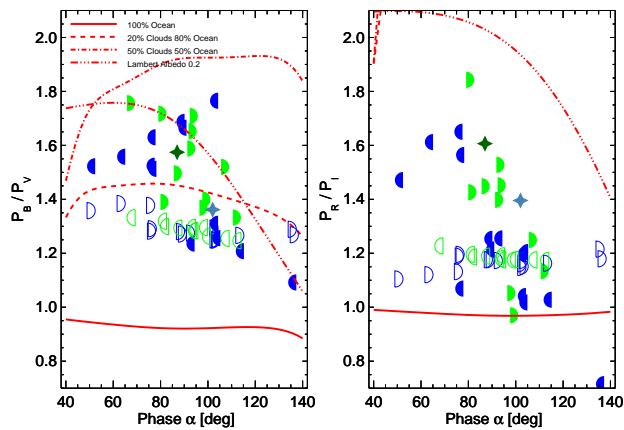}} 
\caption{Polarization color ratios $P_B/P_V$ (left) and $P_R/P_I$ (right) as functions of 
         $\alpha$. Earthshine data are indicated by filled symbols. Blue, left half-moon symbols indicate the Pacific side and
       green, right half-moon symbols the Atlantic side. Values derived from 
         Moonshine spectra are indicated by non-filled (open) symbols for the respective sides of
         the Moon from which the moonlight is scattered. 
         Models of \citet{Stam:2008ij} are overplotted with different linestyles. 
         Values derived from model spectra of \citet{Emde:2017ee} are indicated by stars.  }
\label{Fig:AllPRatios}
\end{figure}

In Fig.~\ref{Fig:AllPRatios}, we show the polarization ratios $P_B/P_V$ and $P_R/P_I$ for our Earthshine polarization data as functions
of phase angle $\alpha$, using different (filled) symbols  for the observations covering the Pacific side
and the Atlantic side. Errors on these ratios are within a few percent, and thus of the order 
of the symbol sizes. The polarization ratios for the data that is corrected for the lunar depolarization
are indeed similar to the data that are uncorrected, thus confirming our choice for using 
polarization ratios.
Figure~\ref{Fig:AllPRatios} also shows model computations from \citet{Stam:2008ij} and \citet{Emde:2017ee}. 
Spectra with a relative shallow decrease of polarization with increasing wavelength have relatively 
small polarization color ratios, while flat spectra have $P_B/P_V=1$ and $P_R/P_I=1$.
For both color ratios $P_B/P_V$ and $P_R/P_I$, there is a general tendency for an anti-correlation with phase angle $\alpha$: 
the smaller $\alpha$, the steeper the polarization spectra. 
However, for $\alpha >$ 100\degree \; 
all spectra appear to be flatter, and a few spectra from the "P" sample even exhibit slightly 
increasing slopes in the red spectral range.  

It is interesting to compare the Earthshine (resp. polarization efficiency corrected) polarization 
color ratios with those for Moonshine. As explained above, each Earthshine observation is associated 
with a Moonshine observation, which is captured in the part of the detector that sees a region of 
sky adjacent to the lunar limb and that is used for background subtraction. The Moonshine consists
of scattered moonlight, and can be used to extract polarization spectra in the same way as we did for 
the Earthshine data. 
In Fig.~\ref{Fig:AllPRatios} we included the Moonshine polarization color ratios, indicated with 
non-filled symbols. Over a wide range of phase angles, the slopes of Moonshine spectra, and thus 
their polarization color ratios, are distinct from those of the Earthshine/Earth spectra. 
While the absolute value of the local lunar polarization depends significantly on the local lunar 
albedo, there is no significant difference in the color ratios of the Moonshine originating 
from the west- or the east-side of the Moon. For the phase angle range considered here, the 
polarization color ratios of the Moonshine appear to be rather independent of phase angle. 
This behavior has been noted before by e.g. \citet{1964AJ.....69..826G} and \citet{1992Icar...99..468S}. 

The polarization color ratios of the Earthshine can also be compared to model simulations. 
As before, we use a representative set of Earth polarization models from \citet{Stam:2008ij}. Conceptually,
the simplest case is a model composed of a Rayleigh-scattering atmosphere with an ocean surface below.
The ocean albedo is zero for all wavelengths, and while at the shortest wavelengths the atmosphere is 
optically thick enough for multiple scattering that lowers the polarization, with increasing wavelength, 
the polarization increases to its single scattering value of nearly 0.9 (the Fresnel reflection lowers
the polarization slightly when compared to a black Lambertian surface), and becomes largely 
wavelength independent.
This mechanism is largely independent of $\alpha$, thus the expected polarization color ratios 
are close to one for all phase angles, as can be seen in Fig.~\ref{Fig:AllPRatios}. 
A Lambertian reflecting surface with an albedo of 0.2 steepens the polarization spectra considerably,
as can be seen from the polarization ratios, as with increasing wavelength, more light will reach the
surface and add more unpolarized light to the Earthshine \citep[see][]{Stam:2008ij}.
The anti-correlation of the measured polarization color ratios with the phase angle is actually 
approximated by such a model for the blue color ratio $P_B/P_V$, but in particular for the red 
color ratio $P_R/P_I$ the model is much too steep, as can be seen in Fig.~\ref{Fig:AllPRatios}.

The various dependencies of the polarization color ratios on wavelength and phase are even 
more complicated when clouds are considered. It is well known that clouds with their rich 
and complex macro- and microphysical properties have manifold impact on polarization spectra. 
With a 20\% cloud coverage fraction in the models of \citet{Stam:2008ij} (combined with 80\%
clear ocean surface), the polarization color
ratios increase to values qualitatively compatible with our observations for the blue ratio 
$P_B/P_V$, but they are off-scale for the red ratios $P_R/P_I$.
Increasing the cloud coverage to a more realistic 50\% (with 50\% clear ocean surface) 
steepens the polarization spectra even more, and appears to become incompatible with 
both the observed blue and the red polarization ratios. 
For comparison, we also include the polarization color ratios derived from the simulations 
of \citet{Emde:2017ee} to fit datasets ID A.3 and B.4. As expected, their more detailed 
cloud and surface properties, as derived from remote sensing data at the time of the 
observations, appears to flatten the spectra and lower the polarization ratios, 
and agree more quantitatively with the observations, but does not match them within the errors.

\subsection{Polarization Vegetation Index}\label{PVI}

\begin{figure} 
\resizebox{\hsize}{!}{\includegraphics[width=11cm, trim = 1.5cm 1.5cm 0 0]{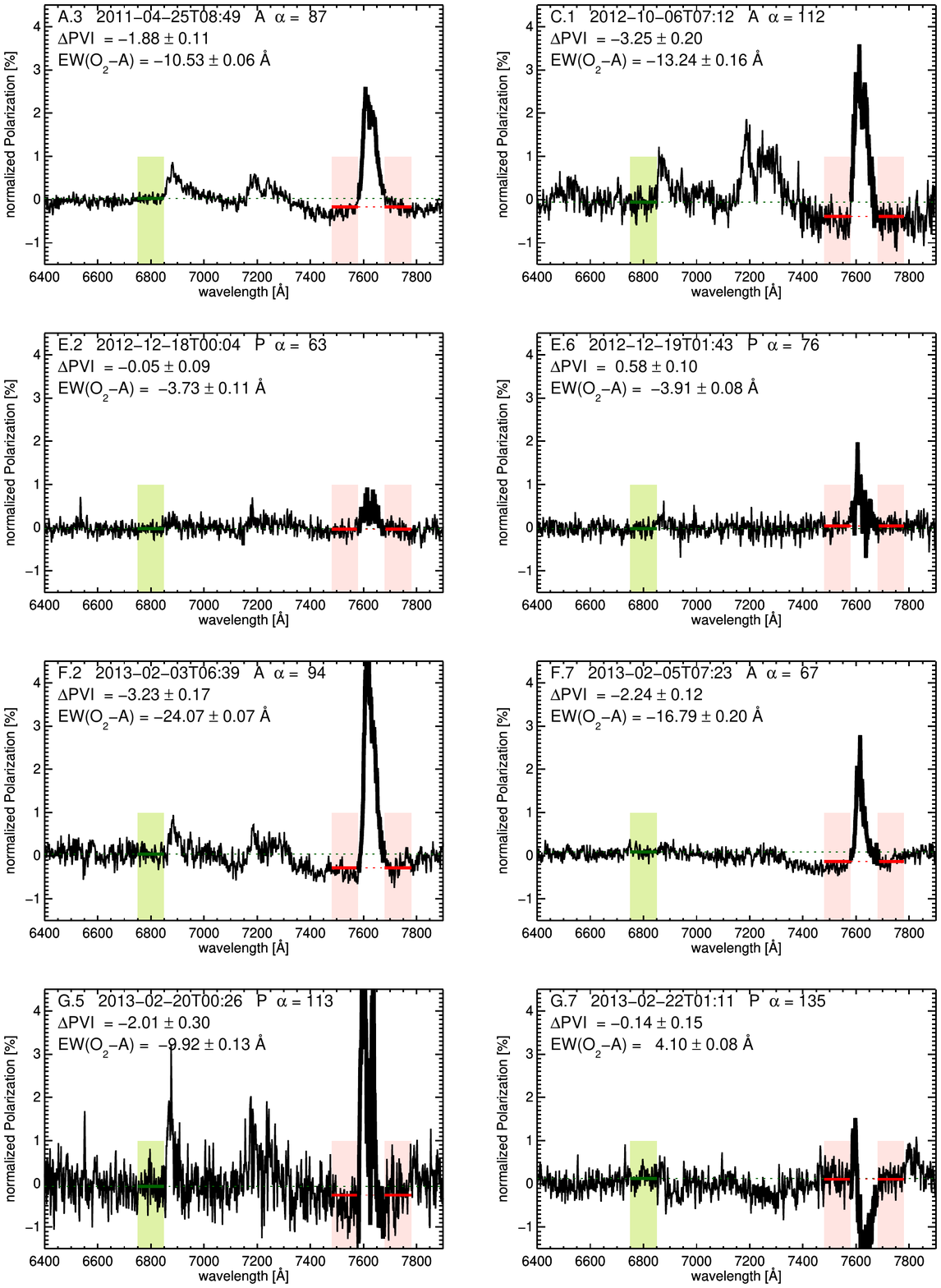}} 
\caption{A selection of normalized spectra between 6400~\AA \; and 8000~\AA. The horizontally 
         dotted lines indicate the regions across which the vegetation index $\Delta$PVI, 
         associated with the Vegetation Red Edge:
         blue-ward (between 6750~\AA \; and 6850~\AA; green dotted line) and red-ward (between 7480~\AA \; and 
         7780~\AA, but avoiding the region affected by the O$_2$-A between 7580~\AA \; and 
         7680~\AA; red dotted line). The difference between the green and red dotted lines defines the amount of $\Delta$PVI. The derivation of EW(O$_2$-A) is explained in the text.}
\label{Fig:DPVI}
\end{figure}

The previous section concentrated on the shape of the continuum polarization spectra.
In order to study gaseous absorption bands and other variations in the spectra, 
we normalize the spectra by subtracting a continuum, following the same procedure described 
in \citet{Sterzik:2012gk}: a fourth order polynomial is fitted to a spectral range between 
5300~\AA \; and 8900~\AA. Spectral regions that contain band and/or variations above 1.5~sigma 
above the mean values are excluded from the fit. The resulting fit is subtracted from the 
original spectrum, and the residuals then represent the spectral variations on the spectrum.
Figure~\ref{Fig:DPVI} displays a representative set of observations in the interesting 
spectral region between 6400~\AA \; and 8000~\AA. This region contains the molecular 
absorption bands O$_2$-B (around 6900~\AA), H$_2$O (around 7200~\AA), and O$_2$-A (around
7600~\AA), as well as the Vegetation Red Edge, the spectral region where the reflectivity
of Earth's vegetation sharply increases \citep{Tinetti:2006bk}.

The Vegetation Red Edge shows up in the polarization spectra mainly because of the added unpolarized
flux that is reflected by the vegetation.
In order to have a closer look at the Vegetation Red Edge, we eliminated effects of the molecular
absorption bands as much as possible. We therefore defined two bands, blue-wards and red-wards 
of the vegetation red edge \citep{Tinetti:2006bk} in the polarization spectra: between 6750~\AA \; 
and 6850~\AA \, and between 7480~\AA \; and 7780~\AA, but avoiding the region affected by the 
O$_2$-A band between 7580~\AA \; and 7680~\AA. 

\subsubsection{Analysis of $\Delta$PVI}

We averaged the normalized polarization over all 
wavelengths across both regions, and call that difference the "Differential Polarization 
Vegetation Index" ($\Delta$PVI). Errors of the $\Delta$PVI are defined by the standard 
deviation in the bands. Negative values of $\Delta$PVI 
indicate a lower continuum in the blue than in the red as a result of a sharp increase of the 
albedo of vegetated surfaces in the red.

\begin{figure} 
\resizebox{\hsize}{!}{\includegraphics{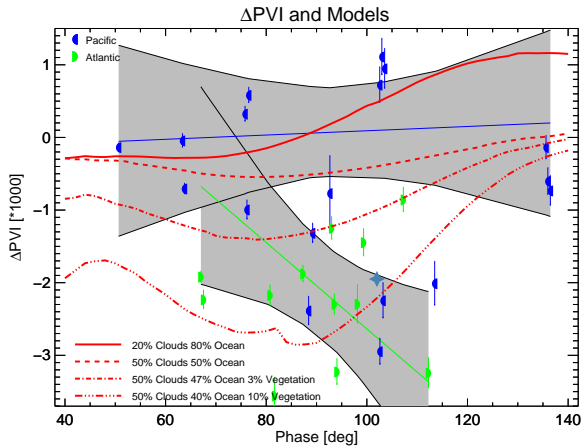}} 
\caption{$\Delta$PVI as a function of the phase angle with the symbols indicating the viewing 
         sceneries (the Pacific or the Atlantic). The green and blue lines indicate the linear 
         regression fit from all the observations of either scenery.  One-$\sigma$ confidence bands around the linear regression have been
         shaded in grey for the two sub-samples.
         While the Pacific side 
         shows a very weak correlation with the phase angle, the population from the Atlantic exhibits decreasing $\Delta$PVI with increasing phase angles. Both samples are statistically distinct by a low probability (0.14\%) in a two-sample KS-test.  Red lines indicate the $\Delta$PVI's
         as derived from the models from \citet{Stam:2008ij}. The star-symbol indicates the 
         $\Delta$PVI as derived from the model for the B.4 from \citet{Emde:2017ee} 
         (the value derived from the model for the A.3 dataset is -4.49, and falls outside
         the plot).}
\label{Fig:DPVI_stats}
\end{figure}

Figure~\ref{Fig:DPVI_stats} shows all values of $\Delta$PVI from the observations listed in 
Tab.~\ref{Tab:PEarth} as functions of the phase angle $\alpha$. 
As before, different symbols indicate different sceneries (blue, left-half circles indicate 
the Pacific and green, right-half circles the Atlantic). 

As can be seen in the figure, values of $\Delta$PVI are scattered, and there is only a weak 
correlation with phase angle. But the values of $\Delta$PVI for the Atlantic-side tend to be 
more negative than those for the Pacific-side. We quantify the correlation by a formal linear 
regression analysis of the parameters and their errors. The population is sparse, with outliers 
and errors. Therefore we apply maximum-likelihood estimator techniques described 
in \citet{Kelly:2007bv}. The 1-$\sigma$ confidence bands around the linear regression have been
shaded in grey for the two sub-samples. Both populations become distinct  with increasing significance for larger phase angles. 
While for small phase angles both distributions overlap, the means of the two distribution are distinct by more than 2$\sigma$ for phase angles around 110\degree.
As outliers exist in both populations that are compatible 
with the mean values of the other population, the significance of the different linear regressions is lowered.  
However, a formal KS-test gives a probability of only 0.143\% for the two populations "A" and "P" being drawn from the same underlying $\Delta$PVI distribution. 
In this sense, both distributions are statistically different.

By design, $\Delta$PVI is supposed to be sensitive to the amount of visible surface 
vegetation and caused by the steep albedo change across these wavelength bands. 
In order to compare the observations with models,  we derived 
the $\Delta$PVI parameter using polarization model spectra of \citet{Stam:2008ij} (the
models consider only 1 type of vegetation, i.e.\ deciduous forest). In order to avoid methodological 
biases, we apply the same 
procedure to remove the continuum and to derive  $\Delta$PVI as done for the observations. 
It is evident from the curves referring to the models in Fig.~\ref{Fig:DPVI_stats} that
$\Delta$PVI depends sensitively on the amount of visible vegetation at most phase angles. 
As an ensemble, the observations of the Atlantic side are bracketed by models containing up to 
3\%--10\% surface vegetation. Observations of the Pacific have a larger scatter, but models without vegetation better describe their distribution.

We have included in Fig.~\ref{Fig:DPVI_stats}  the values derived from the two simulations of \citet{Emde:2017ee} as listed in Tab.~\ref{Tab:PEarth}. Their models contain 3\% and 10\% visible vegetation, and the value of $\Delta$PVI=-4.49\textperthousand  \; corresponding to the high vegetation case for observation A.3  falls outside the range in the figure displayed.
 
\subsubsection{Relation to NDVI}

The parameter $\Delta$PVI is sensitive to
the difference in polarization which is supposed to be induced by the abrupt increase of the surface albedo of Earth's 
vegetation in the near infrared. 

The observed range of 
$\Delta$PVI values for the "A" and "P" sub-samples can be explained with models that contain different fractions of surface 
covered by vegetation. According to these models, already small fractions of vegetation will enhance the signal. 
Our measurements appear to be sensitive to vegetation fractions larger than 3\% and supports the earlier claim in 
\citet{Sterzik:2012gk} that "A" and "P" sceneries can be distinguished by their different 
amount of visible vegetation contained in Earthshine. 

The presence of outliers in both ("A" and "P") sub-samples of the $\Delta$PVI distribution likely washes out the statistically differences of the ensemble averages, in particular for smaller phase angles.
As the amount of visible vegetation cover appears to sensitively change the $\Delta$PVI parameter, individual observations are expected to depend sensitively on the actual presence of vegetation in a given underlying scenery. 

We approximate the actual (or true) vegetation cover of individual observation by deriving the "Normalized Differential Vegetation Index" (NDVI). Based on the visible parts of Earth, and their MODIS surface albedo for the R and I band, we calculate $NDVI=(I_{858nm}-I_{645nm})/(I_{858nm}+I_{645nm})$ to measure the proportion of each image pixel covered with vegetation (cf Fig.\ref{Fig:8Earth}) and derive an average vegetation cover of the total illuminated area for each observing epoch.

\begin{figure} 
\resizebox{\hsize}{!}{\includegraphics{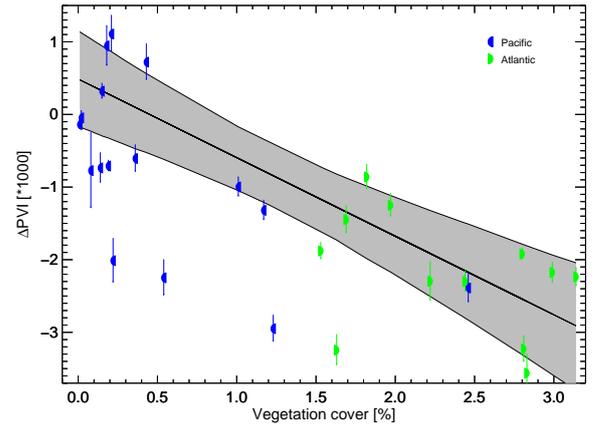}} 
\caption{$\Delta$PVI as a function of the actual vegetation cover observed.
The actual vegetation cover has been calculated from NDVI values for  each image pixel. 
Different symbols indicate the viewing sceneries (the Pacific or the Atlantic). The black line shows a linear 
regression fit with all the observations.  One-$\sigma$ confidence bands around the linear regression have been shaded in grey.
The anti-correlation of $\Delta$PVI with the actual vegetation cover is strong, although outliers exist. 
}
\label{Fig:NDVI}
\end{figure}

Fig.\ref{Fig:NDVI} shows the relation of $\Delta$PVI with the actual vegetation cover observed. 
In general, observations with a  larger actual vegetation cover fraction have a lower value of $\Delta$PVI.
The quantities are anti-correlated, but outliers exist. 
It is interesting to note that even the "P" sample sometimes contains observed sceneries with a  significant vegetation cover, and vice versa. 
$\Delta$PVI appears to be sensitive to detect these cases. 

We conclude the discussion of potential effects of vegetation on our data by noting that the series of observations (B.1 to B.5) contained a broad "bump" in their polarization spectra around 5000~\AA \;  (see Fig.~\ref{Fig:AllSpectra}). We cautioned earlier that we cannot 100\% exclude artifacts caused by the flat fielding procedure. Interestingly,  B.x data are the only ones observing the Pacific ocean during Northern summer. 
We speculate that this bump could also be due to surface reflection: if the albedo increases towards 5000~\AA \; (green vegetation), P will decrease there. 
A possible absorber - that might only be present seasonally on the ocean - could be algae (chlorophyll B).
However, the efficiency of this mechanism needs to be modeled before any further conclusion can be drawn. 

\subsection{O$_2$-A band strength}\label{O2A}

The spectral resolution of our polarization spectra also allows us to determine the strength of 
the polarization signal in specific absorption band regions, most prominently in the O$_2$-A band
region (around 7600~\AA). The polarization in this band is often higher than in the adjacent continuum,
sometimes it is flat, and rarely lower than in the continuum. The variability of the polarization 
in the band can readily be seen for the sample of spectra shown in Fig.~\ref{Fig:DPVI}.
Various processes determine the polarization in an absorption band as compared to that in the continuum: 
the absorption of light by e.g.\ O$_2$ decreases the amount of multiple scattered light, with usually
a lower degree of polarization than the singly scattered light. This process will yield a band
polarization higher than that in the continuum. Absorption of light by O$_2$ will also limit
the amount of light that is scattered at low altitudes in the atmosphere. Indeed, the stronger the
absorption, the lower the altitude from which the Earthshine originates. In a vertically
inhomogeneous atmosphere, different types of particles at different altitudes can yield a  
polarization that varies across the band. In particular, the cloud top altitude will influence
the band strength. Finally, with increasing gaseous absorption, 
less light that has been reflected by the surface will reach the top of the atmosphere. 
Because a reflecting surface will usually increase the amount of unpolarized flux, increasing
absorption will increase the polarization of the Earthshine.
For more detailed explanations and sample computations for whole planet signals, 
see \citet{Fauchez:2017kx}.

To quantify the behavior of this feature we introduced a quantity called "equivalent width" ($EW$), which is frequently used in stellar spectroscopy: here, it measures the area of polarization ($P_\lambda$) over wavelength ($\lambda$) integrated over a specific spectral region (from $ \lambda_0$ to $\lambda_1$) normalized to its continuum value $P_{\rm c}$:  
\begin{equation}
EW(\lambda) = \int_{\lambda_0}^{\lambda_1} (1-P_\lambda/P_{\rm c})  \, d\lambda  
\end{equation}
in practice, we numerically integrate $P_\lambda$ across the passband $\lambda_0$=7580~\AA \; to $\lambda_1$=7680~\AA \; 
\st{normalized} divided by the adjacent continuum level $P_{\rm c}$. We determined $P_{\rm c}$ by a second order 
polynomial fit of two 1000~\AA \; wide regions red- and blue-wards of the band. Negative values 
of $EW$ indicate a band in "emission". We estimate the error in $EW$ and determine its values 
also for both regions in the continuum. These values indicate the intrinsic error of the 
integration over a flat spectral region not affected by the spectral band. The high quality of 
the polarization spectra, and the very good sampling of the O$_2$-A band region lead to small 
errors for the determination of $EW$, typically less than 1\%. $EW$ is largely independent 
of the spectral resolution used, and its values obtained by measurements with different grisms 
can be compared directly.

\begin{figure} 
\resizebox{\hsize}{!}{\includegraphics{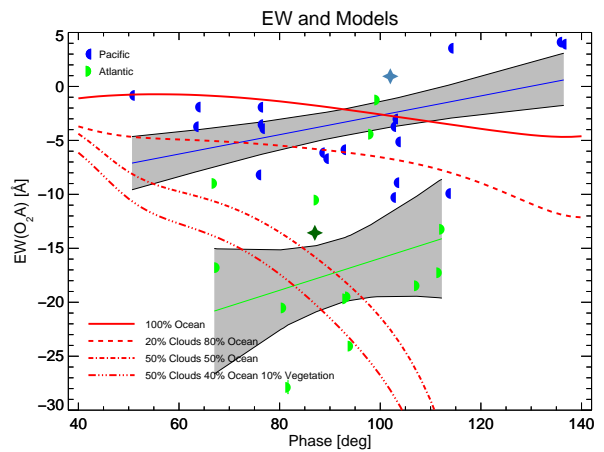}} 
\caption{The equivalent width ($EW$) in the O$_2$-A band as a function of phase angle $\alpha$. 
         The symbols indicate the different sceneries (Pacific and Atlantic). Full lines and 
         grey areas indicate the linear regression and its errors. Both regression curves are distinct by $\approx 2\sigma$ from each other, 
         and offset over the full range of phase angles covered. A two-sample KS-test gives a very low probability ($2.4\cdot10^{-5}$) 
         that both samples are drawn from the same underlying population. Red lines with different 
         linestyles refer to models of \citet{Stam:2008ij} and star symbols refer to the 
         two models of \citet{Emde:2017ee}.}
\label{Fig:EWO2A_stat}
\end{figure}

Figure~\ref{Fig:EWO2A_stat} shows the values of $EW$(O$_2$-A) for all the observations listed in 
Tab.~\ref{Tab:PEarth} as functions of the phase angle. Like in Fig.~\ref{Fig:DPVI_stats}, different 
symbols indicate different sceneries of the Earth.
Evidently, there is a considerable scatter of the $EW$s, and only a marginal correlation 
within the range of phase angles for the Pacific and Atlantic sides. 
Again, we apply a robust least-square fitting procedure to both samples as explained above, 
and show the two regression lines in Fig.~\ref{Fig:EWO2A_stat} for the blue (Pacific) and 
green (Atlantic) sets. The grey areas correspond to a $\pm1\sigma$ confidence interval around 
the optimal regression curves.  Both regression curves are distinct by $\approx 2\sigma$ from each other, 
and offset over the full range of phase angles covered. As in the case of $\Delta$PVI, outliers exist for both samples that are compatible with the other populations. 
A formal two-sample KS-test gives a very low probability ($2.4\cdot10^{-5}$) that both samples are drawn from the same underlying population.

In order to increase our understanding of the behavior of the $EW$(O$_2$-A), we have extracted 
the same quantity for a set of models of \citet{Stam:2008ij}. Although these model spectra are 
at a significantly lower spectral resolution than our observations, they do allow deriving the 
$EW$ in the same way. Results for different model planets (all with the same cloud top altitude)
are plotted in Fig.~\ref{Fig:EWO2A_stat}
with different linestyles.
Qualitatively, the models suggest an anti-correlation between the $EW$(O$_2$-A) and the
phase angle with increasing cloud coverage fraction. 
Interestingly, cloud coverage fractions 
larger than 50\% introduce significant variations in the $EW$(O$_2$-A) values. 

We have also determined the $EW$(O$_2$-A) for the two simulations of \citet{Emde:2017ee} (see Tab.~\ref{Tab:PEarth}). They show that the introduction of cloud layers at different altitudes 
and with different optical thicknesses and droplet sizes have significant effects on the 
appearance and strength of the O$_2$-A band in polarization spectra. For a suitable choice 
of cloud parameters it may thus not be surprising that the $EW$(O$_2$-A) derived from their
models correspond better to the observations.

\begin{figure} 
\resizebox{\hsize}{!}{\includegraphics[angle=0]{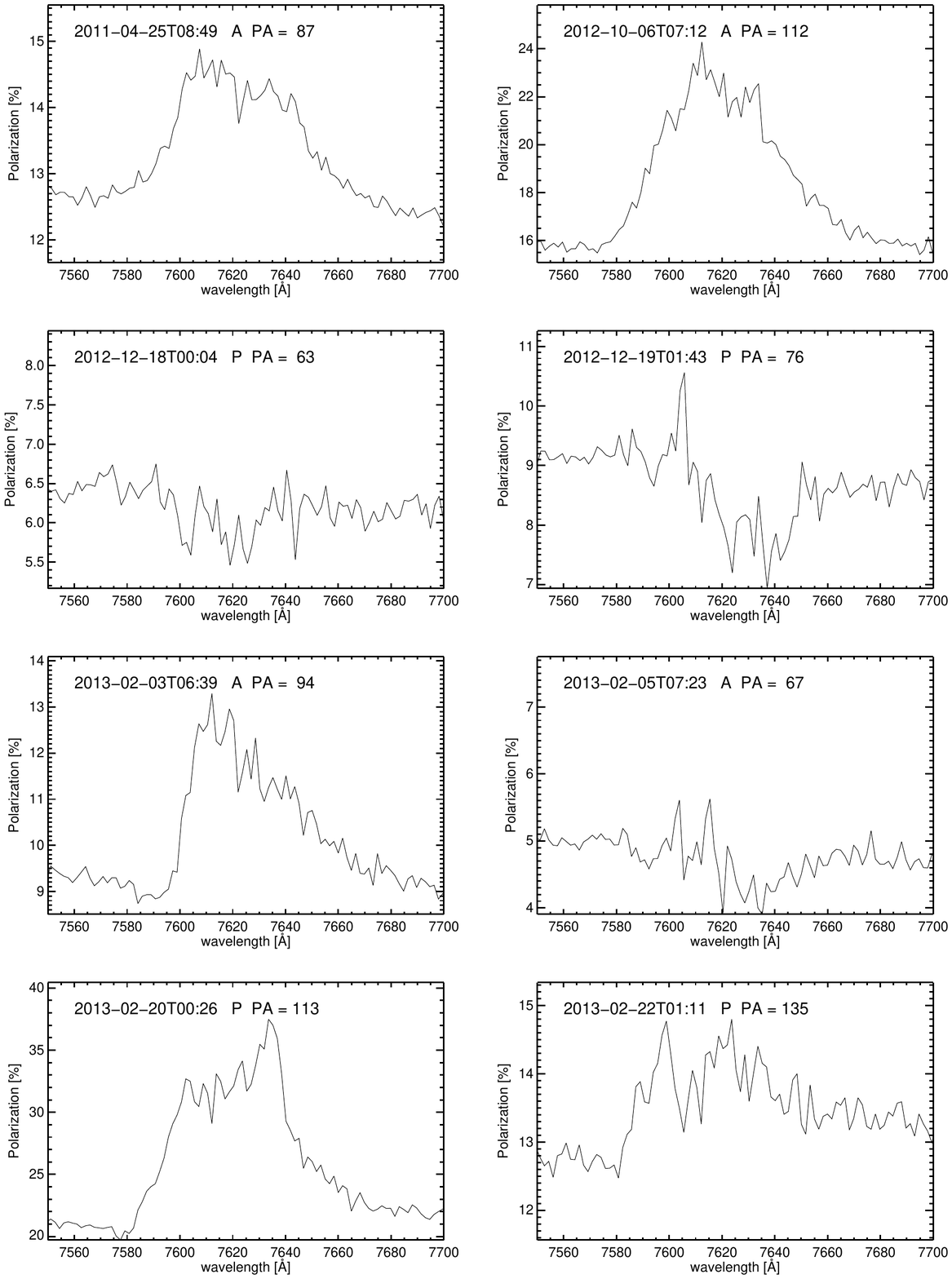}} 
\caption{O$_2$-A band region between 7550~\AA \; and 7700~\AA \; for 8 high-spectral resolution 
         polarization spectra observed with grism 600I. Spectra have been corrected for lunar 
         depolarization, i.e. they should correspond to Earth's true polarization values at 
         these epochs. The fine structure seen in the absorption band is real.}
\label{Fig:O2A}
\end{figure}

Next we focus on observations of the O$_2$-A band obtained with higher spectral resolution.
The appearance of this band in polarization depends on the fraction of (usually highly polarized)
single scattered light to that of the (usually low polarized) multiple scattered light, which 
increases with increasing absorption. It also depends on the vertical distribution of scattering
particles, including cloud particles, in the atmosphere, and it depends on the surface albedo.
Examples of this band in polarization can be seen in Fig.~\ref{Fig:O2A} for those Earthshine spectra 
observed using grism 600I. Resolved fine structure in the band is clearly visible, with some 
parts of the band showing polarization to more than 15\% above the continuum value of 20\%, while others 
do not show enhanced polarization, or even slightly reduced polarization. 
Increasing the spectral resolution would further enhance the contrast in the band
\citep[see][for examples]{Stam:1999fa}. 
More examples with high spectral resolution have been shown in \citet{Emde:2017ee}. 
The new observational contribution in this work, however, is the large 
variation of this line independent of phase and not directly correlated to "A" or "P" sceneries. 
\citet{Fauchez:2017kx} investigate the influence of the planet surface albedo, cloud optical 
thickness, altitude of the cloud deck and the O$_2$ mixing ratio on the polarization in the 
O$_2$-A band, and find that these parameters may not be easily discriminated in the higher optical 
depth regimes that can be probed with high spectral resolution. 
The cloud deck altitude and horizontal distribution across the region of the Earth that 
contributes strongest to the observed signal may be decisive factors for the appearance
of the band, but more detailed simulations of Earth polarization measurements are needed 
to confirm this.
 
\subsection{Short-term variability}\label{STV}

In this section we try to disentangle the three main effects that have an impact on $P^{\rm E}$ on a timescale of a few hours: \\
(a) continuous phase angle changes due to the movement of Earth and Moon, \\
(b) Earth rotates and makes different parts visible at different times, \\
(c) changes in cloudiness are introduced by changing weather patterns. 

As described in Sect.~\ref{DA} and in Tab.~\ref{Tab:Log}, polarimetric spectra were usually acquired with 16 different settings of the retarder waveplates. 
Four 
settings are sufficient to reliably derive the Stokes parameters $Q, U$, and thus $P$, albeit with correspondingly less $S/N$.  
Therefore we can increase the temporal resolution of the  observations to about 15 minutes (the typical duration of an observing cycle with 4 retarder settings). The spectra are then equally processed as described in Sect.~\ref{Sect_Observations} and  \ref{Sect_Data_Reduction}, and the same parameters  (with their statistical and systematic errors) determined as described in Sect.~\ref{Results}.

In particular the datasets observed on 2012-12-19, 2013-02-03, 2013-02-19 and 2013-02-22 allow a continuous monitoring of the polarization of Earth over about 3 hours, with the sampling time of 15 minutes as defined by a full observation cycle. We use these data to investigate the variation of $P^E$ in the four passbands, and of $\Delta$PVI and $EW$(O$_2$-A) on this timescale. 

\begin{figure} 
\resizebox{\hsize}{!}{\includegraphics{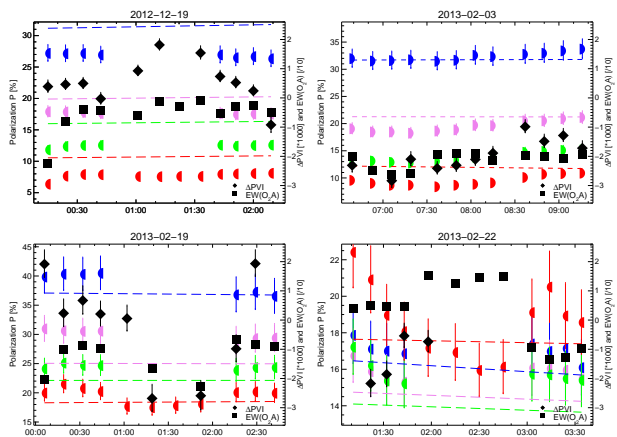}} 
\caption{Short-term variability of polarization features of selected observations in timescales of a few hours. Polarization in the $B,V,R,I$ bands is shown in the usual colors (blue, violet, green, red). $EW$(O$_2$-A) and  $\Delta$PVI  values are plotted with black squares and diamonds in a common scale indicated on the right axis.}
\label{Fig:PolTime}
\end{figure}

Fig.~\ref{Fig:PolTime} displays the variation of these parameters over time. The four panels correspond to 4 distinct observations dates. The colored, half-moon symbols, refer, as before, to the values of $P^E$ defined for passbands $B, V, R, I$. The errors indicated correspond to the (dominant) systematic errors associated with the uncertainty of the lunar albedo, corrected in the same way as discussed in Sect.~\ref{depol}. Dashed lines indicate the global fit solution using the empirically determined parameters from Tab.~\ref{Tab:Phasefits} for the modified Rayleigh function (Eq.~\ref{eq:rayleigh}) of the phase function
 (cp. Fig.~\ref{Fig:Earthpol}). In Fig.~\ref{Fig:PolTime} we show only the subsets corresponding to the exact time (resp. phase angle) when the data-sets were observed. Typically, the phase angle $\alpha$  changes  within 3 hours by not more than 1\degree, and the change of $P^E$ expected for this change in $\alpha$ is less then 1\%. The straight dashed
lines indicate the expected change of $P^E$ {\sl only due to changes of $\alpha$.}  This change is rather low within three hours. 
Within the errors, many of the measured values of   $P^{\rm E}$ in fact are fully compatible with a slow change expected (see eg. datasets 2012-12-19 and 2013-02-03). The measurements are not expected to  follow the empirical fit exactly. The fits correspond to an average sampling of Earth's global appearance, and the offsets may thus just hint to deviations of the actually observed scenery from the averaged, global, one. Datasets 2013-02-19, and in particular 2013-02-22, show more variation, in particular in the $I$ band. For both cases, the intermediate measurements were observed with another grism (600I) than the ones preceding and following (300V). While an offset that is at least partially caused by systematic effects due to the different instrument setup is not excluded, larger amplitude trends and variations are also notedin the other bands, in particular around 01:30UT.

In the same Fig.~\ref{Fig:PolTime} we also plot the values of $\Delta$PVI and $EW$(O$_2$-A) with different symbols, together with their formal errors, in a common scale. 
Apparently, both parameters show some variability,  but are relatively constant in particular for  2012-12-19 and 2013-02-03. Measurements of 2013-02-03 correspond to the "A" sample, and $\Delta$PVI  is lower compared to the others.  Within datasets 2013-02-19, $\Delta$PVI shows excursions to rather low $\Delta$PVI values, indicative for the temporal appearance of a free surface covered by vegetation around UT01:30. It needs to be confirmed if this interpretation holds, and a corresponding surface or scenery came in sight. In the same dataset we also noted variation of $EW$(O$_2$-A) of around 10~\AA, fluctuating during the duration of the observation sequence. This dataset might be in particular affected by changing sceneries in combination with changing cloud patterns during rotation of Earth.

\section{Conclusion}\label{Discussion}

Earthshine observed with the VLT naturally allows to divide the sample into two groups: one group contains Earthshine from waxing Earth, and contains major contributions from the Atlantic ocean, the Amazonas region, Europe, Africa 
and the Antarctica
(here called sample "A"). 
The other group contains Earthshine for the waning Earth and probes in particular the Pacific ocean, with no or little visible
land surface (sample "P"). 
Both observational sceneries are partially covered by variable, and possibly 
systematically different, cloud patches. They represent different views of planet Earth.
Our statistical analysis focused on finding observational properties that may distinguish the two groups.
This should constrain the impact of typical surface and atmosphere characteristics on observables 
extracted from polarization spectra  and build up an empirical basis for comparison with 
theoretical scattering models of Earth-like (exo)planets. 

We have extracted the fractional polarization in four characteristic wavelength bands  from individual Earthshine spectra 
and constructed Earth phase curves in relation to their phase angle $\alpha$ in Sect.~\ref{Phasecurve}.
Overall, polarization at a certain phase angle is consistent with the values reported in the literature.
But there is considerable spread among datasets from different authors, and among different observing runs of our own.
We observe variability of polarization spectra on timescales ranging from minutes to data which are separated by weeks to months. 

We reduced the largest systematic uncertainty and derive the Earth polarization phase curve after correcting the observed Earthshine data for  lunar depolarization efficiency. Notwithstanding, fluctuation around a mean phase curve, which can be well approximated by a modified Rayleigh function over the phase angle region covered by our data, exists at a relative level of 20\% around the maximum degree of polarization (i.e. $\pm$6\% at $P_{max}$=36\%). This spread can - at least for the blue spectral band - be bracketed by models of Earth having 
clouds covering 40\% of its surface and the other 60\% covered by ocean. Generic theoretical polarization models cannot fully explain the red spectra range, in particular at larger phase angles where $P$ tends to be relatively high, and spectrally flat. 

We have introduced polarization color ratios in the blue, $P_B/P_V$, and in 
the red, $P_R/P_I$ which largely reduce uncertainties caused by lunar depolarization. 
Comparing them with simple models demonstrates their sensitivity to cloud 
and surface properties. 
Using cloud properties from Earth remote-sensing data at the time of the observations, the 
Earth models from \citet{Emde:2017ee} appear to be fully compatible with two observations in the blue
color ration $P_B/P_V$. In the red, there is less scattering by the gaseous atmosphere, and
thus more light will reach the surface. 
The color ratio in the red, $P_R/P_I$ appears difficult to explain with the current models. 
The explanation of the relatively shallow slopes of Earth's polarization spectra in the red,
appears to require more advanced surface reflection in the models.

The difference between the "A" and "P" sub-samples with respect to their different polarization
phase functions, and relatively higher $P_{max}$ values (see Fig.~\ref{Fig:Earthpol}, 
and Tab.~\ref{Tab:Phasefits}) for the "P" sample may indicate a slightly lower average 
cloud coverage when sampling this hemisphere. But also differences of the mean cloud 
optical thickness and/or altitudes can contribute to systematic differences in the polarization
phase functions. Detailed models combined with Earth remote-sensing data tailored to 
individual observations should help to discover the correct explanation.

Reliable phase curves of planet Earth are sparse, respectively absent. 
So far, only few polarization values in three bands $B, R, I$ have been derived from the satellite-borne POLDER instrument by \citet{2005ASPC..343..211W} for a phase angle of 90\degree, extrapolating to whole Earth cloud coverages. Their values for an assumed 55\% global cloud coverage are 22.6\%($P_B$), 8.6\%($P_R$) and 7.3\%($P_I$), lower than those derived by us, and lower than the models by \citet{Stam:2008ij}. 
But we note the potential advantages of monitoring Earth's polarization directly from suitable satellites, which allows cross-calibration between Earthshine and direct satellite polarization data, and should help to further validate theoretical models. Because a satellite in a low Earth 
orbit does not provide an instantaneous view of the whole Earth, instead sampling small regions
of the Earth along its orbit, a suitable satellite would preferable be far away, such as
in a geo-stationary orbit or even in a lunar orbit 
\citep[for a description of the advantages of the latter position, see][]{2012P&SS...74..202K,
2016OExpr..2421435H}.

The insufficient phase coverage of our data did not  allow to detect signatures of enhanced polarization due to rainbows and/or cloud-bows at a scale of global Earth. For water clouds, the primary rainbow is expected for scattering angles around 139\degree, thus at a phase angle around 41\degree, with a small spectral dependence.  \citet{2007AsBio...7..320B} and \citet{Karalidi:2012fc} predict a factor of two to three enhanced polarization around that phase angle, even with cloud coverage fractions as low as 30\%. Unfortunately, the lowest phase probed by our observations was 50\degree, where no
enhanced polarization would be expected \citep{Karalidi:2012fc}.  
It will be interesting to probe those small phase angles in the future, albeit the difficult 
viewing geometry (the dark region on the moon across which the Earthshine would be observed
is indeed very narrow at those phase angles) implies very short observing times for Earthshine. 

Signatures of the glint, i.e. direct sunlight reflected on the (visible) ocean surface can, in principle, occur at most viewing geometries and phase angles, and may be easier detectable in Earthshine than rainbows. \citet{Williams:2008ds}, \citet{Zugger:2010dx} and \citet{Robinson:2010gn} model glint properties on distant ocean-covered planets. Enhanced light curve modulation and/or somewhat higher degrees of polarization signals are predicted. Detection of sunglint on the Earth \st{from} by the satellite LCROSS was reported in \citet{Robinson:2014er}, adding specular reflection on Earth's ocean surface to the inventory of potential diagnostics for a habitable world.  
\citet{Emde:2017ee} 
simulate polarization spectra at phase angles of 87\degree \;
and 102\degree \; corresponding to the observations presented in \citep{Sterzik:2012gk}. 
As planetary surfaces they included a two-dimensional Lambertian
surface and an ocean surface, respectively. Their results show that the
degree of polarization is up to 10\% higher for the ocean surface and
they attribute this change to the highly polarized sun-glint.
In relative terms, the increase of $P$ in the red part of the spectrum is largest, up to 10\%. 
The spread observed in our data may at least be partially affected by variable conditions for glint polarization and we refer to detailed modeling in a forthcoming paper. 

We found that the actual vegetation coverage for a given observing scenery, as measured through NDVI, correlates 
with the $\Delta$PVI parameter derived from the polarization spectra around bandpasses characteristic for the Vegetation Red Edge.  
$\Delta$PVI thus allows to probe contributions of surface vegetation on Earth rather sensitively and robustly. 
The positive detection 
of a VRE -- even at different wavelengths -- had been proposed as a potential biosignature 
for other planets \citep{2006ApJ...644L.129T}, but its interpretation as biosignature appears 
to be problematic as certain types of mineral reflectance could mimic the albedo slope and strength 
\citep{ 2005AsBio...5..372S}. Further characterization and potential application of VRE polarization detection is matter of active research
\citep[see, e.g.][]{Sterzik:2010vc, Martin:2010dh, Berdyugina:2015jn, 2015IAUS..305..346B, Martin:2016gc}.

Understanding the O$_2$-A (and B) absorption band is essential to substantiate its 
relevance as potential biomarker in exoplanet atmospheres. 
\citet{2014ApJ...781...54R} and \citet{Lovis:2017kt} assess the feasibility to detect this band 
for the closest exoplanets with current and future high-resolution (flux) spectrographs at the 
VLT and the ELT. Our observations stress the potential of spectropolarimetry as a double-differential 
technique to enhance contrast between the (unpolarized) host star and the highly polarized 
planet exhibiting a large dynamics in the fractional polarization across this line.
Another advantage of using spectropolarimetry to detect O$_2$ absorption in an exoplanetary
atmosphere is that 
the polarization across the band is independent of the absorption by O$_2$ in the 
Earth's atmosphere (except that the latter absorption will lower the observable fluxes
and the signal to noise ratio).
Future feasibility studies should be directed to assess the merit employing 
high-resolution spectropolarimetry to observe this band in exoplanets.
 
\begin{acknowledgements}
Based on observations collected at the European Southern Observatory under ESO programmes P87.C-0040 and P90.C-0096. We thank the comments of an anonymous referee which helped to improve the paper.
\end{acknowledgements}

\bibliographystyle{aa}
\bibliography{allpapers}

\end{document}